\documentclass[a4paper,12pt]{article}
\usepackage{graphicx,natbib,fullpage}
\usepackage{setspace}
\usepackage[noblocks]{authblk}
\usepackage{amssymb}

\def\aj{AJ}
\def\apj{ApJ}
\def\aap{A\&A}
\def\aaps{A\&AS}
\def\procspie{Proceedings of the SPIE}
\def\apjs{Astrophysical Journal Supplement Series}
 
\title{A Two-Colour CCD Survey of the North Celestial Cap:\\I. The Method}
\author[1]{Evgeny Gorbikov}
\author[1]{Noah Brosch}
\affil[1]{\small{The Wise Observatory and
the Raymond and Beverly Sackler School of Physics and Astronomy,
the Faculty of Exact Sciences,
Tel Aviv University, Tel Aviv 69978, Israel}}
\author[2]{Cristina Afonso}
\affil[2]{\small{Max Planck Institute for Astronomy, K\"{o}nigstuhl 17, 69117 Heidelberg, Germany}}

\begin{document}
\onehalfspacing
\maketitle
\begin{abstract}
We describe technical aspects of an astrometric and photometric survey of the North Celestial Cap (NCC), 
from the Pole ($\delta=90^{\circ}$) to $\delta=80^{\circ}$, in support of the TAUVEX mission. 
This region, at galactic latitudes from $\sim17^{\circ}$ to $\sim37^{\circ}$, has poor coverage in modern CCD-based surveys. 
The observations are performed with the Wise Observatory one-meter reflector and with a new mosaic CCD camera (LAIWO) 
that images in the Johnson-Cousins R and I bands a one-square-degree field with sub-arcsec pixels. 
The images are treated using IRAF and SExtractor to produce a final catalogue of sources. The astrometry, based on the USNO-A2.0 
catalogue, is good to $\sim$ 1 arcsec and the photometry is good to $\sim$0.1 mag for point sources brighter than R=20.0 or I=19.1 mag. 
The limiting magnitudes of the survey, defined at photometric errors smaller than 0.15 mag, are 20.6 mag (R) and 19.6 (I). 
We separate stars from non-stellar objects based on the object shapes in the R and I bands, attempting to reproduce the SDSS 
star/galaxy dichotomy. The completeness test indicates that the catalogue is complete to the limiting magnitudes.
\end{abstract}

\section{Introduction}
The North Celestial Cap Survey (hereafter NCCS) is compiled from digital imaging observations of 
the Northern Celestial Cap region ($\delta \geq +80^{\circ}$) 
performed at the Wise Observatory from February 13, 2009 with the 1-meter Boller and Chivens telescope
and  the Large Area Imager at Wise Observatory (LAIWO) digital camera. The NCCS is a photometric 
and astrometric catalogue in the Johnson-Cousins R and I bands \citep{JOH53,COU74}
and is expected to contain more than 1,500,000 distinct objects.
This paper presents technical details of the project and a brief discussion of the quality of the first results.

The NCCS catalogue of point and extended objects 
supports the TAUVEX project.  The TAUVEX space telescope array, constructed by ElOp
(Electro-Optic Industries Ltd.), a division of ELBIT Systems, for Tel Aviv University with funding
 from the Israel Space Agency (Ministry of Science, Culture, and Sport), consists of a
bore-sighted assembly of three 20-cm telescopes imaging in
 the vacuum UV the same $\sim$one-degree field of view from geosynchronous orbit. Satellites in such orbits and, in particular, those used for telecommunications, 
are normally not used for astronomy since they do not point-and-track celestial objects. To allow the observation of various 
objects in the sky TAUVEX is mounted on the side of
the Indian Space Research Organization (ISRO) GSAT-4 satellite on a Mounting Deck Plate (MDP) 
that can aim the TAUVEX line of sight (LOS) to different declinations, from
$\delta$=+90$^{\circ}$ to $\delta$=-90$^{\circ}$. As GSAT-4 orbits
the Earth on its geo-synchronous orbit, the TAUVEX LOS
scans a sky ribbon. The data transmitted to the ground station are
reconstructed into a set of UV images of the sky ribbon scanned by
the experiment. 

The scanning mode of observation used by TAUVEX implies that the motion of objects through the TAUVEX field 
of view is done at the sidereal rate. Because of this, the exposure times for each source vary with declination
 $\delta$ as $\frac{1}{cos\delta}$. In order to reach very deep exposures without requiring numerous 
re-scans of the same sky ribbon, the TAUVEX observations will mostly be restricted to the circumpolar regions and, 
for the first half-year of the mission, the area to be observed will mainly be 90$^{\circ}\leq |\delta| \leq 80^{\circ}$;
the northern sky patch covered by NCCS.

TAUVEX in survey mode uses its three principal filters SF-1, SF-2
and SF-3. These filters span the spectral region from somewhat
longer than Lyman $\alpha$ to 320 nm with three well-defined
bands. Three filters define two color indices in the UV that can be combined with optical (e.g., V-R or R-I) 
and eventually infrared color indices to characterize the nature of detected sources.

The availability of visual and near-infrared photometric digital sky surveys from 1990's, such as DSS, SDSS and 2MASS, made the process 
of retrieving astronomical data easy as never before and contributed greatly to the development 
of modern astronomy. \cite{RES05} presented a very good review of sky surveys and deep fields. Yet, 
very few high-quality photometric data are available for the Northern Celestial Cap region. The NCCS 
aims to produce a catalogue of positions and R-I color indices to complement the UV colours obtained by 
TAUVEX. 

The first high-quality digital sky survey, the Digitized Sky Survey (DSS), was produced 
by scanning the plates of photographic surveys (POSS-I, POSS-II, ESO/SERC) with specific photometric calibrations. 
Although the DSS and its extension DSS-II are 
both all-sky surveys and cover the North Celestial Pole region, they are based on photographic 
observations and suffer from photographic emulsion shortcomings, such as low sensitivity, 
limited dynamic range and non-linearity. The limiting magnitude of the survey also differs for 
different directions, depending on which 
photographic survey was used to retrieve the data (B$_{lim}\cong20.^{m}0$ for POSS-I, $\cong22.^{m}5$ 
for POSS-II and $\cong18.^{m}5$ for ESO/SERC). 

Another great ccontribution of these photographic surveys (POSS-I, POSS-II, ESO/SERC) 
was to provide high-precision astrometry used in catalogues such as USNO-A1.0, USNO-A2.0, USNO-B1.0, etc.
The USNO-A and USNO-B catalogues were all-sky high-precision astrometric catalogues including also photometric data. 
USNO-A included two-color (B and R) one-epoch data, while the USNO-B catalogue included three-color 
(B, R and I) and two-epoch data.
The USNO-B1.0 catalogue included photometric data with an accuracy of 0.3 mag and astrometric data with 
an accuracy of 0.2 arcsec for more than
 a billion individual objects as faint as V$=21.^{m}0$ \citep{MON03}. The catalogue provided 
separation with 85\% accuracy of stars and non-stellar objects with internal magnitudes $14\leq M_V\leq20$. The USNO-B1.0 catalogue was released at the 
same time that the SDSS early data release became available.
\cite{MON03} compared the two data sets and found ``\textit{systematic offsets as large as 0.25 arcsec
... taken as evidence for distortions of the USNO-B1.0 astrometric calibration}''. Moreover, when transforming 
the catalogue magnitudes to SDSS magnitudes and 
comparing common objects, systematic photometric errors as high as 0.20 mag and dispersions up 
to 0.34 mag were found, which can be explained probably by the photographic nature of the catalogue.

Another digital sky survey is the Two Micron All Sky Survey (2MASS). 2MASS was performed in the 
near-infrared (NIR): J (1.25 $\mu$m), H (1.65 $\mu$m) 
and K$_s$ (2.17 $\mu$m) with two robotic telescopes in the USA and Chile in 1997-2001.
The NIR was chosen to minimize the influence of galactic and extragalactic dust on the photometric data.
2MASS is not suitable as a reference survey for the TAUVEX project by itself, since it is NIR only and is not sufficiently deep, 
with limiting magnitudes J = 15.8, 15.0 mag, H = 15.1, 14.3 mag and K$_s$ = 14.3, 13.5 mag for point 
and extended sources respectively \citep{SKR06}. 

The Sloan Digital Sky Survey (SDSS) covers about 10,000 square degrees 
of the sky and provides high-precision photometric data in five Sloan bands with
limiting magnitudes of $u_{lim} = 22.^{m}0$, $g_{lim} = 22.^{m}2$, $r={lim} = 22.^{m}2$, $i_{lim} = 21.^{m}3$,
$z_{lim} = 20.^{m}5$ for point objects and an astrometric accuracy of $<0.''1$ RMS per coordinate \citep{ABA09}.
SDSS is not suitable as well for the TAUVEX project purposes, since it does not cover the North Celestial Cap, but 
will be used here for comparison, as a commonly accepted standard.

The NCCS is essential for the TAUVEX project since no available survey completely satisfies 
the requirements to support TAUVEX in this sky region. Here we describe the method, the achieved 
accuracy, and provide a comparison with SDSS.
In a following paper we will present some preliminary results.

\section{Observations}

The observations were performed at the Wise Observatory, Israel, from February 13, 2009 in runs of 3-9 nights per month. 
The 1-meter telescope 
and the LAIWO camera were used. LAIWO is a mosaic CCD camera built at the Max Planck Institute 
for Astronomy, Heidelberg.
The telescope and camera parameters are listed in Table \ref{Tab:param}.

\begin{table}[htbp]
\caption{The Telescope and Camera Parameters:}
\begin{center}
\begin{tabular}{l|l}
\hline
\multicolumn{1}{c}{\textbf{Parameter}} & \multicolumn{1}{|c}{\textbf{Value}} \\ \hline \hline
\multicolumn{ 2}{c}{\textbf{Telescope}} \\ \hline
Clear aperture & 101.6 cm \\ 
Focal length (f/7) & 711.2 cm \\ 
Secondary mirror diameter & 51 cm \\ 
Field of view & $\sim$4.9 square degrees \\ \hline
\multicolumn{ 2}{c}{\textbf{LAIWO}} \\ \hline
Number of CCDs & 4 science + 1 guider \\ 
Nimber of pixels & 4$\times$4096$\times$4096 (science) \\ 
Pixel size & 15 $\mu$m \\ 
Field of view & $59'\times59'$ \\ 
Sampling & $\sim$0.$''$44 per pixel \\ 
Peak quantum efficiency & $\sim$40\% between 600 and 850 nm \\ 
Read-out noise & $9<$RON$<19$ e$^-$ \\
 & (quadrant-dependent) \\ 
Full-well depth & $\sim$80,000 e$^-$ \\ 
Gain & 5 e$^{-}$ ADU$^{-1}$ \\ \hline
\end{tabular}
\end{center}
\label{Tab:param}
\end{table}

The technical parameters of LAIWO determined during its construction were described by \cite{BAU06}. 
For a more detail description of LAIWO and its performance see Afonso et al. 2010, in
prep. Nevertheless, we dedicate the following subsection to describe its characteristics and performance for a better
understanding of the potential and adequacy of LAIWO for the NCCS survey.

\subsection{LAIWO}

\begin{figure*}[ht!]
\begin{center}
\includegraphics[angle=0,width=140mm]{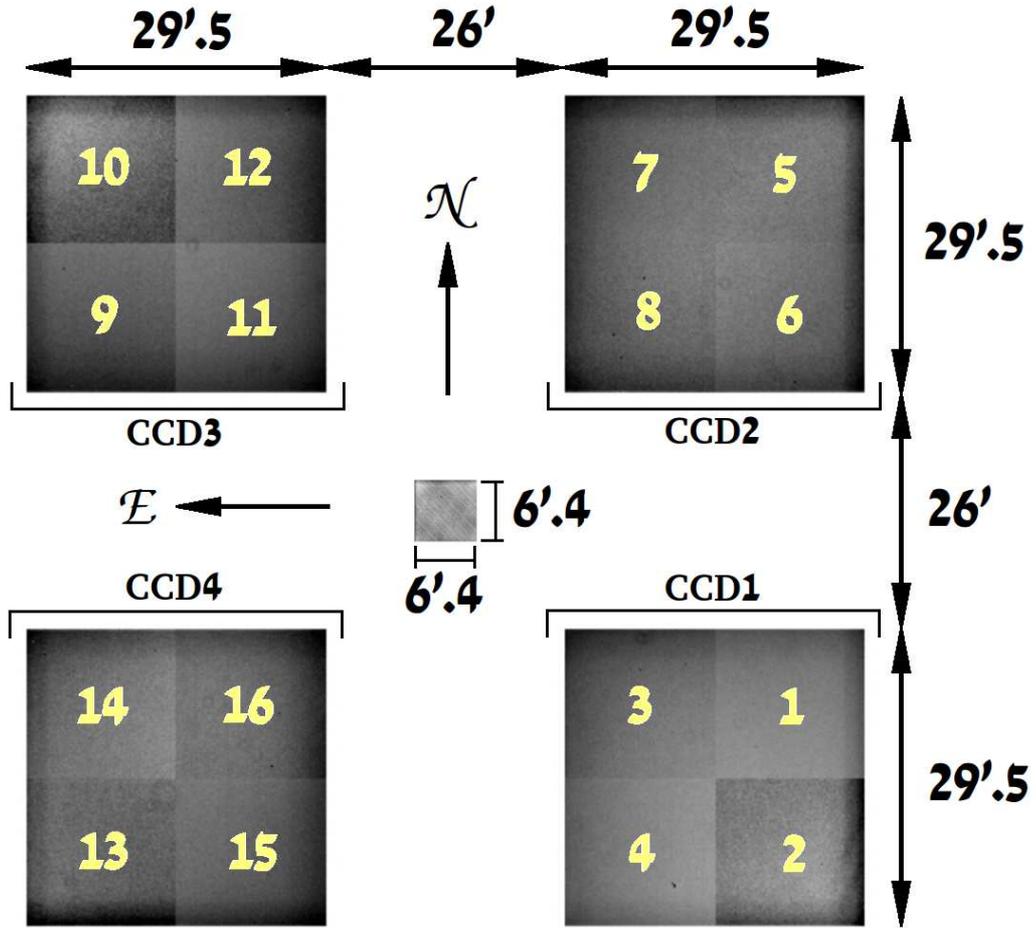}
\end{center}
\caption{
I band flat field imaged with LAIWO science CCDs on June 12, 2009 used to show LAIWO CCD layout. 
A clear-filter flat field imaged with LAIWO guider CCD on July 5, 2009 is shown at the center of the image.
The relative sizes and distances are preserved. 
   \label{Fig:layout}}
\end{figure*}

The LAIWO camera consists of four Lockheed CCD486 4096$\times$4096 pixel frontside-illuminated devices,
refered to as the science CCDs. The LAIWO observation manual and some technical details can be found in \cite{KAS09} 
and \cite{AFO07}. 
At the f/7 focus of the Wise telescope each pixel subtends 0.43 arcsec and each CCD images a 
29.5$\times$29.5 arcmin$^2$ field. The CCDs are mounted on a single heat sink cooled by liquid nitrogen (lN$_2$) to 
-105$\pm5^{\circ}$C and are individually connected to the lN$_2$ dewar with flexible copper bands. The chips are not contiguous,
but are spaced $\sim$26 arcmin apart. Each science CCD 
is connected to four output channels to reduce the read-out time of the entire mosaic. 
Figure \ref{Fig:layout} shows the layout of the LAIWO 16 quadrants in North-East orientation. The resultant image 
of an exposure is a mosaic FITS file \citep{HAN01} consisting of 16 extensions. 

The science CCDs 
exposures are binned 2$\times$2 to match the typical seeing, reducing the number of pixels to 16 Mpixel sampled to
$\sim0.''87$ arcsec per binned pixel. The read-out time of the entire mosaic in 2$\times$2 binning is only $\sim$28 sec.

A guider CCD is located at the center of the science CCDs mosaic. This is a back-illuminated e2V CCD47-20 device
with 1024$\times$1024 13$\mu$m pixels, corresponding
to 0.38 arcsec pixel$^{-1}$, with maximal quantum efficiency of $\sim$80\% and covering a $6.'4\times6.'4$ field of view. 
The science and the guider CCDs can 
be exposed separately
and/or simultaneously. The guider CCD usually takes continuous short exposures of a small region including a guiding 
star. The image is compared to the previous one and, if the shift of the star photocenter exceeds certain limits, 
the telescope pointing is corrected via a LAIWO-telescope interface. The guider CCD images are not stored by default at the end 
of the night.

\begin{figure*}[ht!]
\begin{center}
\includegraphics[angle=0,width=120mm]{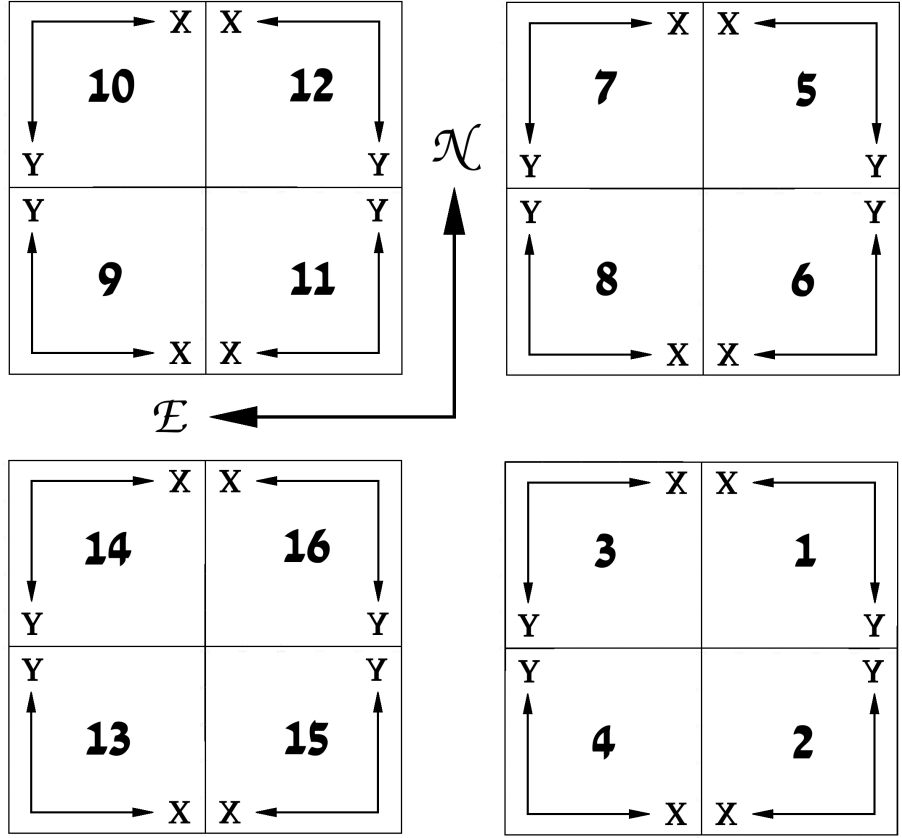}
\end{center}
\caption{
WCS defined for each LAIWO quadrant. The relative sizes and distances are not preserved here. The CCD identification and
orientation are like in Fig. \ref{Fig:layout}.
   \label{Fig:WCS}}
\end{figure*}

LAIWO has no dead or hot pixels, 
however all the flat field images taken with the camera have all the pixels in column X=1 and in row 
Y=1 saturated in each quadrant. Row Y=2 in each quadrant shows sometimes a few saturated pixels. 
Fig. \ref{Fig:WCS} shows the World Coordinate System specific to each LAIWO quadrant. Note that the saturated
columns X=1 and rows Y=1 in each LAIWO quadrant are the edge columns and the edge rows of each CCD;  
this saturation feature is not observed in science images. 
For this reason, sources located near the image edges are not extracted.

The entire LAIWO CCD array detects 1090$\pm$30 cosmic rays (CR) during a 300-sec exposure
($\sim$3.6 CR*sec$^{-1}$). Most CRs produce $\sim$100-300 counts above the
background, with only $\sim$7\% of the CRs produce more than 1,000 counts above the background.
Table \ref{Tab:CR} shows the average numbers of the CRs detected in each CCD from four 300-sec 
exposures of the same field taken on May 14 and June 13, 2009. 

\begin{table}[htbp]
\caption{LAIWO CR Sensitivity:}
\begin{center}
\begin{tabular}{c|c|c}
\hline
 & \multicolumn{ 2}{c}{CR number} \\ \hline
CCD \# & Total & $>$ 1,000 counts  \\ \hline
1 & 280$\pm$20 & 18$\pm$4 \\ 
2 & 270$\pm$20 & 22$\pm$5 \\ 
3 & 270$\pm$20 & 20$\pm$4 \\ 
4 & 270$\pm$20 & 22$\pm$5 \\ \hline
\textbf{total} & 1090$\pm$30 & 80$\pm$10 \\ \hline
\end{tabular}
\end{center}
\label{Tab:CR}
\end{table}

The edge-to-center (EC) flat field (FF) ratio of LAIWO is about the same for the R and I bands and is $\sim86\%$. 
Table \ref{Tab:E2CFF} shows the EC values for each quadrant, extracted from twilight FF images from 
June 14, 2009. The center FF value 
was extracted as the mean value of the 16 pixels of the particular quadrant located near the CCD center. 
The edge value was extracted as the mean value of the 16 pixels furthest from the CCD.

\begin{table}[htbp]
\caption{Edge-to-Center Flat Field Ratio of the LAIWO Camera:}
\begin{center}
\begin{tabular}{c|c|c}
\hline
 & I-band Flat Field & R-band Flat Field \\ 
Quadrant \# & Edge-to-Center Ratio & Edge-to-Center Ratio \\ \hline
1 & 84.9$\pm$0.7\% & 87.9$\pm$0.7\% \\ 
2 & 90.9$\pm$0.7\% & 88.1$\pm$0.8\% \\ 
3 & 86.6$\pm$0.7\% & 86.3$\pm$0.7\% \\ 
4 & 88.3$\pm$0.7\% & 83.9$\pm$0.7\% \\ 
5 & 84.6$\pm$0.7\% & 87.0$\pm$0.7\% \\ 
6 & 84.8$\pm$0.7\% & 83.7$\pm$0.7\% \\ 
7 & 84.8$\pm$0.7\% & 86.1$\pm$0.7\% \\ 
8 & 86.3$\pm$0.7\% & 83.9$\pm$0.7\% \\ 
9 & 84.8$\pm$0.7\% & 86.0$\pm$0.7\% \\ 
10 & 93.8$\pm$0.7\% & 88.9$\pm$0.7\% \\ 
11 & 83.0$\pm$0.7\% & 85.0$\pm$0.7\% \\ 
12 & 87.6$\pm$0.7\% & 85.1$\pm$0.7\% \\ 
13 & 88.8$\pm$0.7\% & 89.3$\pm$0.8\% \\ 
14 & 88.0$\pm$0.7\% & 85.9$\pm$0.7\% \\ 
15 & 85.4$\pm$0.7\% & 85.7$\pm$0.7\% \\ 
16 & 87.3$\pm$0.7\% & 83.5$\pm$0.7\% \\ \hline
\textbf{Global average} & 87$\pm$3\% & 86$\pm$2\% \\ \hline
\end{tabular}
\end{center}
\label{Tab:E2CFF}
\end{table}

\begin{figure*}[ht!]
\begin{center}
\includegraphics[angle=0,width=100mm]{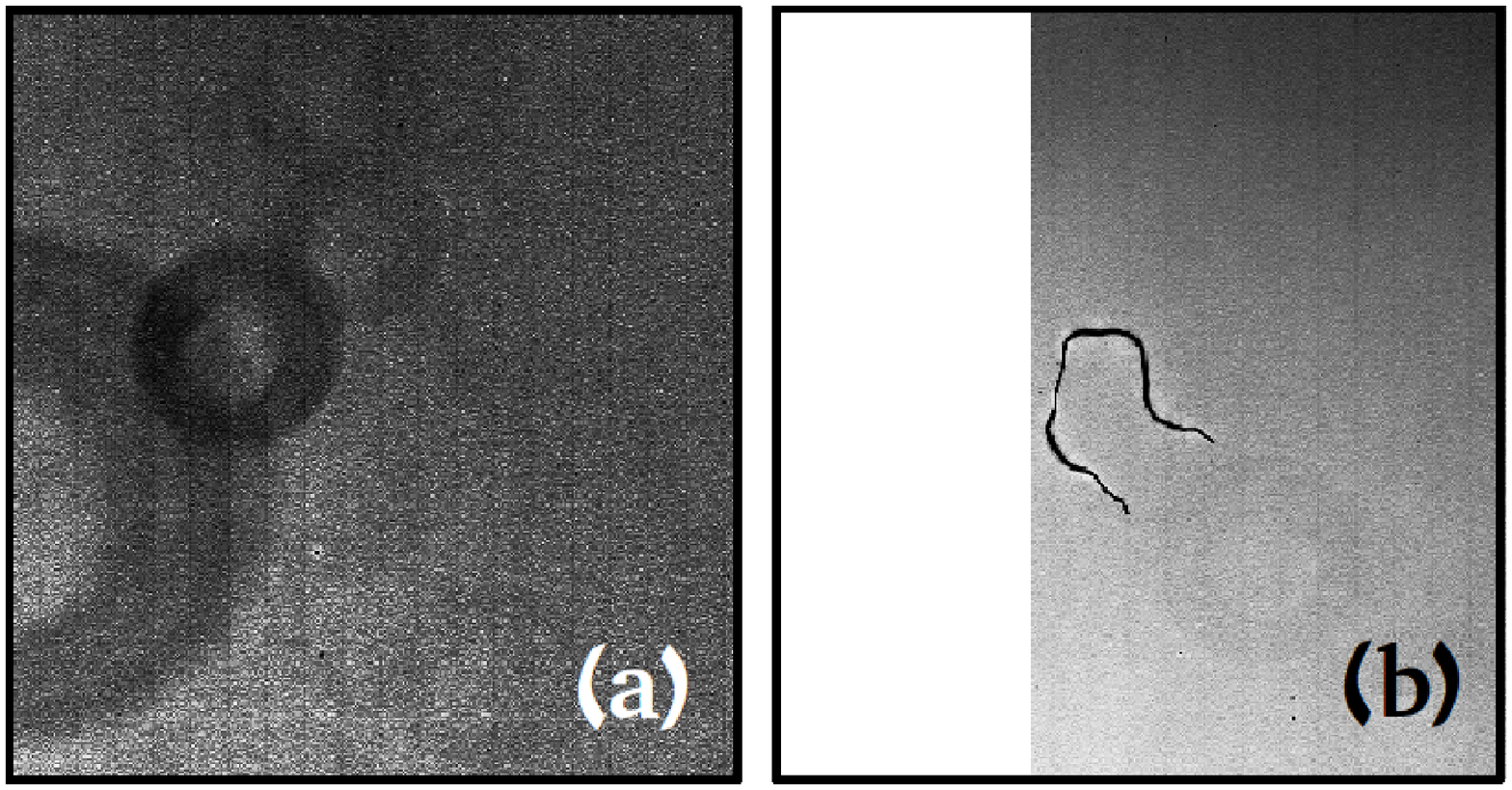}
\end{center}
\caption{
LAIWO dust difraction patterns. Panel (a) shows a donut pattern in I band flat field.
Panel (b) shows a filament pattern in
R band flat field.
   \label{Fig:dust}}
\end{figure*}

The LAIWO camera FFs exhibit different obscuration patterns, which are divided into two groups: 
ring-shaped dust diffraction patterns (``donuts'')
and irregular patterns (``filaments''). Fig. \ref{Fig:dust} shows examples of these patterns, observed on 
FFs from June 12, 2009. The donuts are produced by dust particles located on the filter or on the 
CCD window and affect the exposure level at a $\sim$1\% level. Their location and number may vary from filter to filter
and from night to night. The filaments are produced probably by dust particles or tiny debris located on the CCDs themselves. 
They dim the light by $\sim$20\% and their location is constant for all the filters, 
but their number may change from night to night. These features are observed also in science images. 
We eliminate their influence by using FFs taken in the same night and by imaging the same sky field 
three times with dithering of $\sim$15$''$ , to prevent the appearance of the features on the final, debiased, FF-subtracted
and median-combined image.

A 0-sec exposure of LAIWO camera shows a bias level of $\sim400\pm20$ counts pixel$^{-1}$ (hereafter 1 count = 1 ADU). A 300-sec dark
exposure produces an additional dark level of $3\pm3$ counts pixel$^{-1}$ ($\sim$0.01 counts pixel$^{-1}$ sec$^{-1}$).
Bias/dark exposures taken at the beggining and at the end of a night show the same bias/dark levels. 

Using SExtractor (Source-Extractor, SE; \citealt{BER96}), we obtained \cite{KRO80} aperture parameters 
for all the objects $2\sigma$ above the background noise from an image of the \cite{LAN09} standard field SA107 taken on 
June 14, 2009 at an airmass of 1.175. The Kron aperture is an elliptical aperture fitted individually to every object.
The parameters of Kron aperture are \textit{a}, \textit{b} and $\theta$, the semi-major and semi-minor axes and the position angle
of the main axis, respectively. The \textit{a} and \textit{b} parameters of the Kron aperture are analogs of the FWHM of circular aperture.

\begin{figure*}[ht!]
\begin{center}
\includegraphics[angle=0,width=140mm]{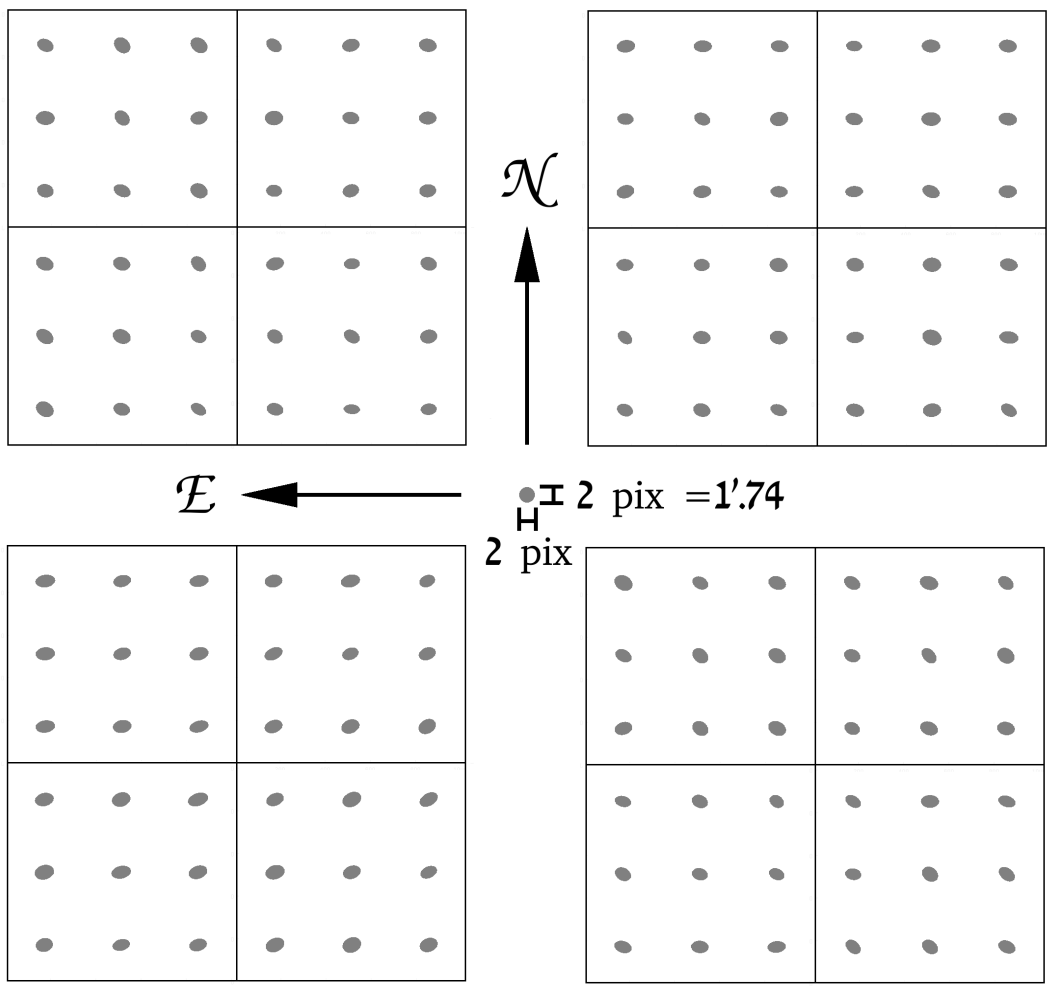}
\end{center}
\caption{
LAIWO distortion picture. The grey circle at the center is for scaling and shows an ideal circular object with
FWHM 2''. The quadrant layout is as shown in \hbox{Fig. \ref{Fig:layout}}.
   \label{Fig:distortion}}
\end{figure*}

By exposing this image at a low airmass we eliminate the effect of atmospheric dispersion which is further reduced
by the filter bandpass restriction. 
The ellipticity of the objects may arise from an optical distortion, from some CCD tilt, from inaccurate tracking/guiding 
or from a combination of these factors. 
We divided every quadrant into 3$\times$3=9 
regions and calculated the mean value of \textit{a}, \textit{b} and $\theta$ of the objects in them. The results 
are plotted in Fig. \ref{Fig:distortion}. In a number of quadrants the major axes are
directed mostly from the camera center outwards. The exception is CCD2 (upper-right corner) where
the ellipses seem to be directed almost randomly. Note that the ellipse elongation on the X axis is greater than that on 
Y axis.
      
\subsection{Observational Strategy}

The NCCS observational strategy is to obtain three 300-sec exposures of each field in the two Johnson-Cousins filters R and I,
with a small $\sim$15$''$ dithering between the exposures. The three 300-sec exposures are then registered and median-combined to allow
the detection of fainter
objects and to reduce random noises such as sky and object Poisson noise, dust interference patterns, 
cosmic rays, satellite tracks, etc. Biases, 300-sec darks and 
twilight FFs are also taken at the beginning of each night. \cite{LAN09} photometric standard stars are observed, usually during one 
night of the run, whenever the weather conditions allow. For runs when no Landolt standards are observed, the photometric 
calibrations are derived from the overlapping regions of the particular run and other runs.

The fields are chosen using two selection criteria. The first is that they should be as close to the meridian as possible and they 
should be higher than the North Celestial Pole to be observed at the lowest possible airmass. We try
to follow the meridian during the night, however the hour angle deviation from the meridian may sometimes be up to $3^{h}$.
The second criterion is that the following field should have a small overlap region with the previous field to provide a
contiguous coverage of the NCC region, filling in the ``empty'' spaces between the CCDs in LAIWO array.  


\section{Data Reduction}

The reductions are done using a fully-automated pipeline written in IRAF (Image Reduction and Analysis Facility; \citealt{TOD86}) script.
The pipeline uses the SE program \citep{BER96} to obtain object fluxes from the images, and the 
WCSTools (World Coordinate Systems Tools)
package of programs \citep{MIN02} to extract astrometric standards from USNO-A2.0 catalogue \citep{MON98}. The pipeline input
consists of all the images
of a particular night. The pipeline output is a list of objects containing the parameters defined in Table \ref{Tab:output}.
Most of the data are extracted automatically by the SE routines. The pipeline runtime 
(including reductions, photometric and astrometric calibrations) is about 6 hours for 
one observing night.

\begin{table}[htbp]
\caption{Pipeline Output Parameters:}
\begin{center}
\begin{tabular}{c|c|c}

\hline
\textbf{\#} & \textbf{Output Parameter} & \textbf{Remarks} \\ \hline
1 & Image date &  \\ 
2 & Filter &  \\ 
3 & Airmass &  \\ 
4 & Quadrant number &  \\ 
5 & Object number &  \\ 
6 & Object position along X &  \\ 
7 & Object position along Y &  \\ 
8 & RA (J2000.0) &  \\ 
9 & DEC (J2000.0) &  \\ 
10 & Corrected isophotal flux &  \\ 
11 & RMS error for ISOCORR flux &  \\ 
12 & Kron flux &  \\ 
13 & RMS error for Kron flux &  \\ 
14 & Petrosian flux &  \\ 
15 & RMS error for Petrosian flux &  \\ 
16 & FWHM &  \\ 
17 & Kron radius &  \\ 
18 & Background level &  \\ 
19 & Semi-major axis & of Kron apperture \\ 
20 & Semi-minor axis & of Kron apperture \\ 
21 & Position angle & of Kron apperture \\ 
22 & Ellipticity & of Kron apperture \\ 
23 & Elongation & of Kron apperture \\ 
24 & SE internal flags &  \\ 
25 & Truncation flag &  \\ \hline
\end{tabular}
\label{Tab:output}
\end{center}
\end{table}

The pipeline median-combines all the bias and dark exposures to obtain master bias and master dark frames. 
The master bias is subtracted from the FFs and the debiased FFs are median-combined with mode scaling 
to obtain master flat frames for the R and I bands. The master flats are normalized, each quadrant to its median. 
The master dark 
is subtracted from the science images and the dark-subtracted science images are then normalized by the master flats. 
The normalized science 
images are then split into 16 separate images, one per CCD quadrant, keeping the common WCS aligned (East to the left,
North up). 

All the images are grouped by the sky imaged field, by the number of the quadrant and by the filter. Each group 
contains three 300-sec debiased, dark-subtracted and flat-fielded exposures of the same quadrant in the same 
field, taken with the same filter. For each image in the group the shifts 
produced by dithering are calculated. The earliest taken image between the three is used as the reference frame. 
The images are then registered to cover exactly the same field. The images in each group are then 
median-combined to obtain one less noisy 300-sec  
exposure, as explained above. The resultant image contains only 
the overlapping region of the three original images; rows and columns out of the overlapping region 
are filled by the median value of the overlapping region.

The combined science images are scanned with SE. We prefer SE photometry to IRAF photometry  
since it is very fast and robust. We define the scanning threshold of the SE to be 2$\sigma$ 
above the noise fluctuations, which corresponds to a point-object detection probability of 95.4\%. 

We use \cite{KRO80} aperture photometry to determine the flux of the objects,
which uses an elliptical adaptive aperture, for a number of reasons:
\begin{enumerate}
 \item LAIWO produces elongated objects even at low airmass as shown in Fig. \ref{Fig:distortion}.
 \item The survey area is relatively close to the horizon (Alt = $20^{\circ} - 40^{\circ}$, airmass $\cong$ 2.0), 
therefore all point sources will
be slightly elongated by the atmospheric dispersion. For point sources, this 
seems preferable  to that of a circular aperture.
 \item Elliptical apertures are natural for extended sources. Moreover, galaxy Kron fluxes can be easily transformed
into Petrosian fluxes \citep{GRA05}, which are used in the SDSS. 
 \item The Kron photometry is flexible, since the aperture is fitted individually to each object. 
 \item The Kron photometry provides additional parameters of the detected objects, 
such as semi-major and semi-minor axes, inclination angle, ellipticity and elongation. These parameters are very useful 
for star/galaxy separation and for field distortion estimation.
\end{enumerate}

The SE subtracts the sky counts from the object counts. We set the background subtraction 
parameters of the SE to local background fit and subtraction. 
For each saturated object, or one that overlaps with another, has poor photometry, is truncated, etc. the SE ascribes an internal flag.
The SE defines as truncated an object that is close to the edge of a quadrant, that may not necessarily be the edge of the 
common region of the three initial 300-sec exposures. The source lists need therefore to be cleared from false 
detections produced by objects located only partly in the common region of the combined image. 
The pipeline ascribes the truncation flag to all the objects located closer than 10 pixels to the edge of the common 
region of the combined images. 

The pipeline also calculates the airmass of the image and the effective airmass of the combined 300-sec exposure, which we define as the
mean airmass of the three initial exposures. The airmass that appears in the output lists is the effective airmass of the combined exposure.

\section{Photometric calibration}

The photometric calibration is relative to \cite{LAN09} standards. Currently the photometric calibration program
is not a part of the pipeline. Some modified parts of the pipeline and an adittional program written in MATLAB are used for 
photometric calibrations. If the standard data for a run are missing, we derive calibration equations 
from objects in the overlapping region between a particular run and another run when standards were observed.

The following calibration equations were obtained from the June 14, 2009 night. The Landolt SA107 field, which contains stars 
with R and I magnitudes of about $10^m - 15^m$ and R - I colours from $-0.^{m}5$ to $+1.^{m}4$, was observed at 11 airmasses 
from 1.17 to 3.16 with 120-sec exposures. The calibration coefficients were obtained from the Kron fluxes of 18 standard stars.
The following calibration equations were derived:
\begin{equation}
\begin{array}{rcl}
R&=&-2.5\log(N_R/t)-(0.155\pm0.006)X+(0.001\pm0.042)(R - I)+(20.952\pm0.023) \\
I&=&-2.5\log(N_I/t)-(0.092\pm0.004)X+(0.349\pm0.029)(R - I)+(19.743\pm0.016)
\end{array}
\label{Eq:calib}
\end{equation}
where $N/t$ is the object flux in ADU sec$^{-1}$ and $X$ is the airmass. 

We define as `grey' magnitudes $m_I$ and $m_R$ the object magnitudes obtained from equations (\ref{Eq:calib}) without 
including the colour term.
We also define the instrumental colour $m_R-m_I$ as the difference between the grey magnitudes of the object in the respective filters.
Using calibration equations 
(\ref{Eq:calib}), the relation between the instrumental and Landolt colours is:
\begin{equation}
\begin{array}{rcl}
R - I &=& 0.742(m_R - m_I)\\
\Delta(R-I)&=&0.742\sqrt{\Delta m_R^2+\Delta m_I^2 +(m_R-m_I)^2(0.042^2+0.029^2)}
\end{array}
\label{Eq:colors}
\end{equation}

  
Equation (\ref{Eq:colors}) allows the transformation of fluxes to magnitudes in two stages, since the R and I images of the 
same field would not necessary be of exactly the same field due to small changes in
telescope pointing and image dithering. The detected objects would not have exactly the same X and Y, and some would possibly miss I 
or R exposures till later stages of the survey. Moreover, one needs to know the true (R - I) colour 
of each object to perform the correction of magnitudes in equation (1) for the colour term, and the object true (R - I) colour
is usually not known beforehead. 
Therefore, the photometric transformation needs to be done in two stages: first, to obtain the grey magnitudes from the 
fluxes, and second, after the astrometric solution for the field is found, to correct the grey magnitudes for the colour terms,
using colours obtained from equation (\ref{Eq:colors}). This yields the true R and I magnitudes of the objects. 

\begin{figure*}[ht!]
\begin{center}
\includegraphics[angle=0,width=120mm]{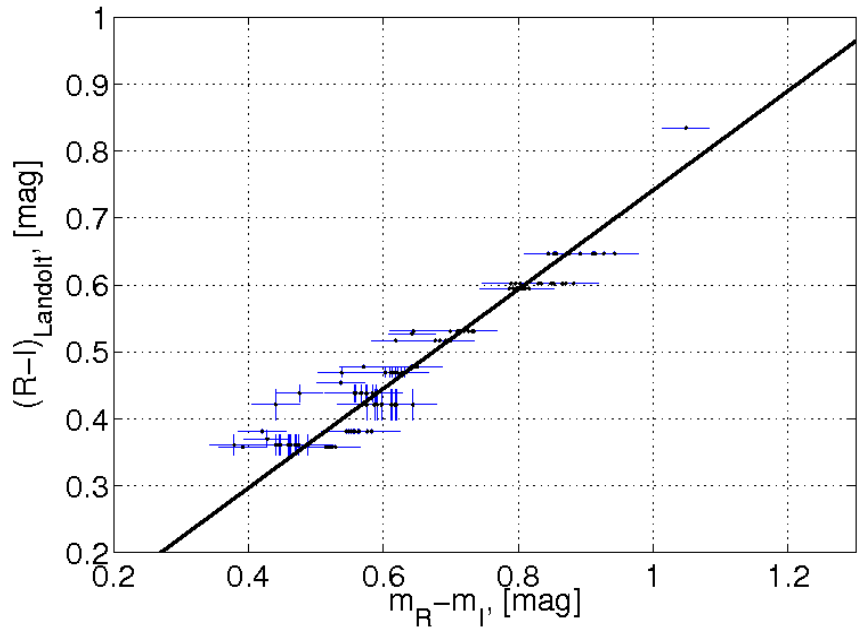}
\end{center}
\caption{
Landolt colour vs. Instrumental colour. The black solid line shows the relation predicted by the equation (\ref{Eq:colors}). 
   \label{Fig:colors}}
\end{figure*}

Fig. \ref{Fig:colors} shows the Landolt colours vs. the calculated instrumental colours and the line predicted 
from equation (\ref{Eq:colors}). The fit of the photometric solution in equation (\ref{Eq:calib}) to the magnitudes ofthe
Landolt standards has a root-mean-square error $\sigma = 0.047$ mag and $\sigma = 0.033$ mag in the R and I bands respectively. 

\section{Astrometric calibration}

Since the telescope pointing is not sufficiently accurate, an astrometric solution needs to be found for every field. 
For the astrometric calibration we use the USNO-A2.0 catalogue, which contains entries for more than half a billion stars with
an accuracy of $\sim$0.25 arcsec and covers the entire sky \citep{MON98}. We use the WCSTools package to extract the needed 
part of the catalogue.
The J2000.0 $(\alpha, \delta)$ coordinates of the extracted part of the catalogue are then transformed to the (X, Y) plane coordinates
using the center of the LAIWO array as a tangential point. The brightest unsaturated stars with $SNR \geq 1,000$ in each LAIWO quadrant
are extracted by the SE and the quadrant output lists are combined into the CCD output lists to obtain a more accurate solution.
The number of the stars used for the astrometric solution in each CCD is usually $\sim$30.

The (X, Y) coordinates of the brightest stars are matched with the (X, Y) coordinates of the extracted catalogue part using 
the downhill simplex algorithm \citep{NEL65}. The algorithm produces an initial match between the coordinate lists with 
an accuracy of $\sim$10-14 pixels. The final astrometric solution is then fitted by IRAF using a tangetial projection 
with parabolic surfaces and a linear distortion along the X and Y axes. The J2000.0 $(\alpha, \delta)$ coordinates of 
each object are calculated and updated in the output lists. The final RMS deviation of the NCCS astrometric data
for the most of fields ranges from 0.5 to 0.8 arcsec and is in all the cases better than 1.25 arcsec
in both RA and DEC. The astrometry was tested also against SDSS; the results are presented in Section 8 below.

\section{Limiting Magnitudes}

\begin{figure*}[ht!]
\begin{center}
\includegraphics[angle=0,width=125mm]{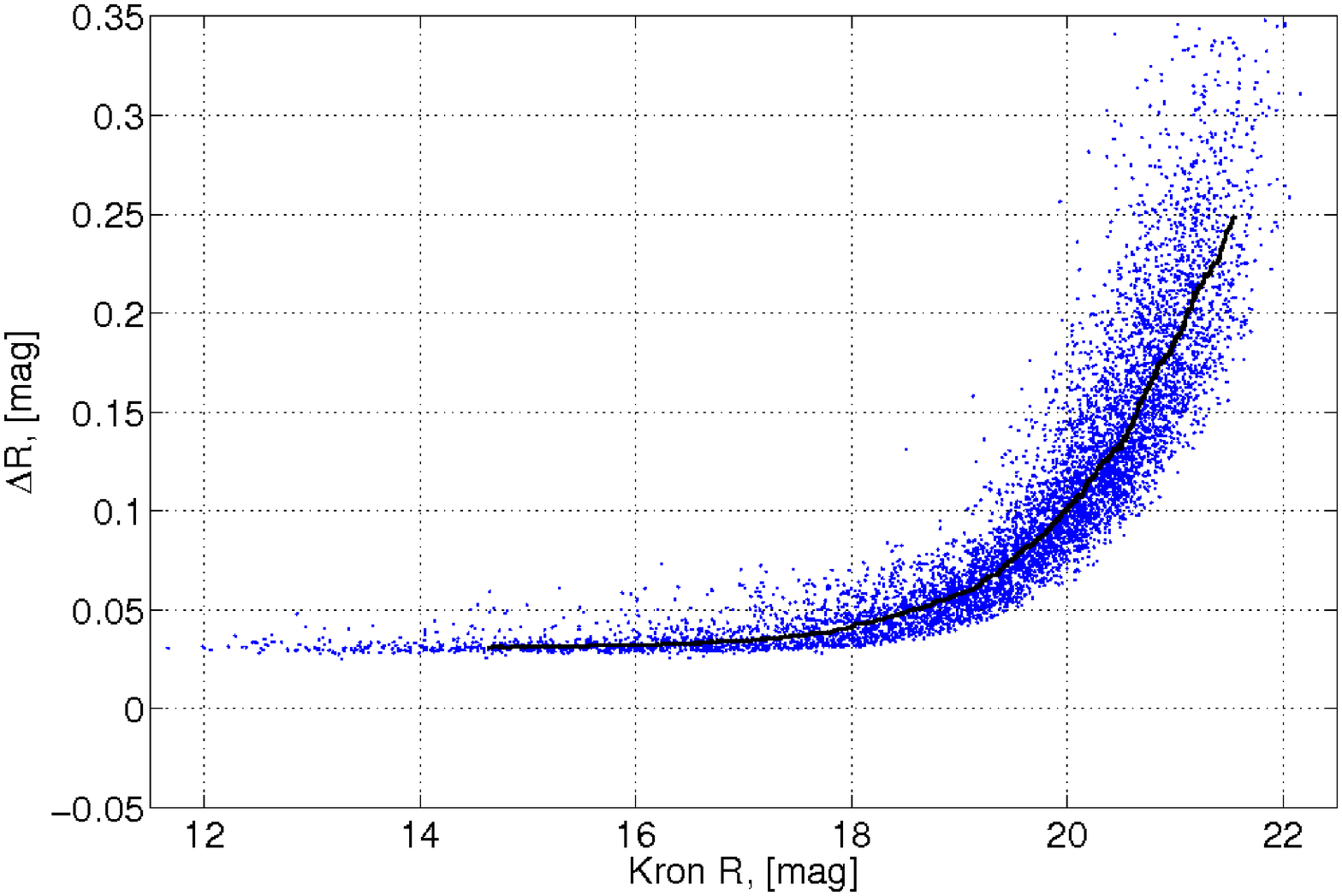}
\end{center}
\caption{Kron R magnitude errors as a function of Kron R magnitudes for $\sim$10,000 objects imaged in the same field. The black
line shows a running median of errors in 400-object bins.  
   \label{Fig:RdR}}
\end{figure*}

\begin{figure*}[ht!]
\begin{center}
\includegraphics[angle=0,width=125mm]{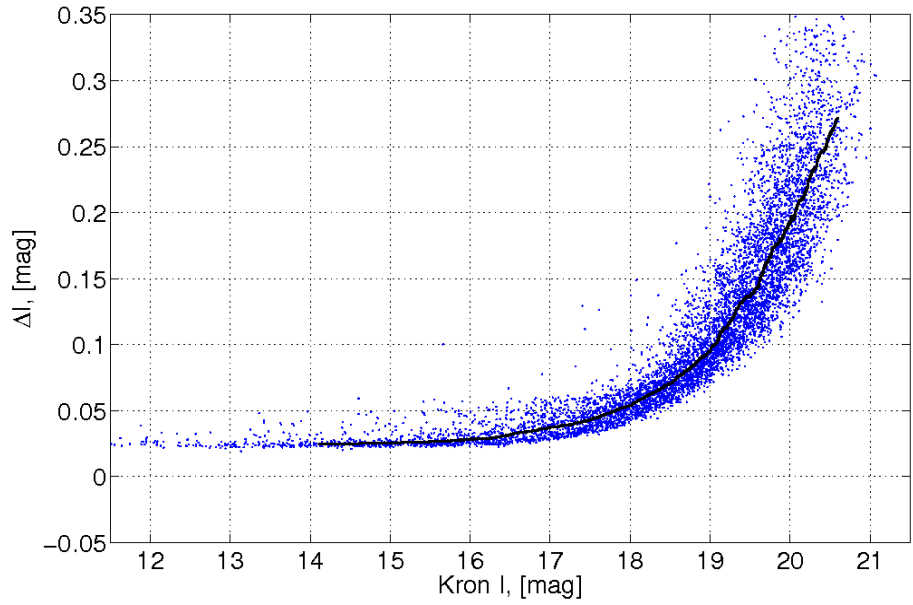}
\end{center}
\caption{Kron I magnitude errors as a function of Kron I magnitudes for $\sim$10,000 objects imaged in the same field. The black
line shows a running median of errors in 400-object bins.  
   \label{Fig:IdI}}
\end{figure*}

Following the derivation of the astrometric solution and the correction of the grey magnitudes for the colour terms, 
the true R and I magnitudes are calculated for each object. Once the transformation operations are performed we can estimate 
the photometric accuracy and depth of our survey.
The saturation limit derived from the images on June 14, 2009 is R = 11.5 mag and I = 11.5 mag for a 300-sec exposure. 
Figs. \ref{Fig:RdR} and \ref{Fig:IdI} show the Kron R and I magnitude errors as a function of Kron R and I magnitudes respectively 
for $\sim$10,000 objects 
from the same field imaged on June 12, 2009. The objects were extracted from a 300-sec combined exposure adopting a minimal 
SNR = 2. Table \ref{Tab:limmag} shows the R and I magnitudes extracted from Figs. \ref{Fig:RdR} and \ref{Fig:IdI} that correspond 
to median errors of 0.05, 0.10 and 0.15 mag.

\begin{table}[htbp]
\caption{R and I Magnitudes Corresponding to the Values of R and I Errors:}
\begin{center}
\begin{tabular}{c|c|c}

\hline
$\Delta$R, $\Delta$I & R & I \\ \hline
0.05 & 18.6 & 17.8 \\ 
0.10 & 20.0 & 19.1 \\ 
0.15 & 20.6 & 19.6 \\ \hline
\end{tabular}
\end{center}
\label{Tab:limmag}
\end{table}

\begin{figure*}[ht!]
\begin{center}
\includegraphics[angle=0,width=125mm]{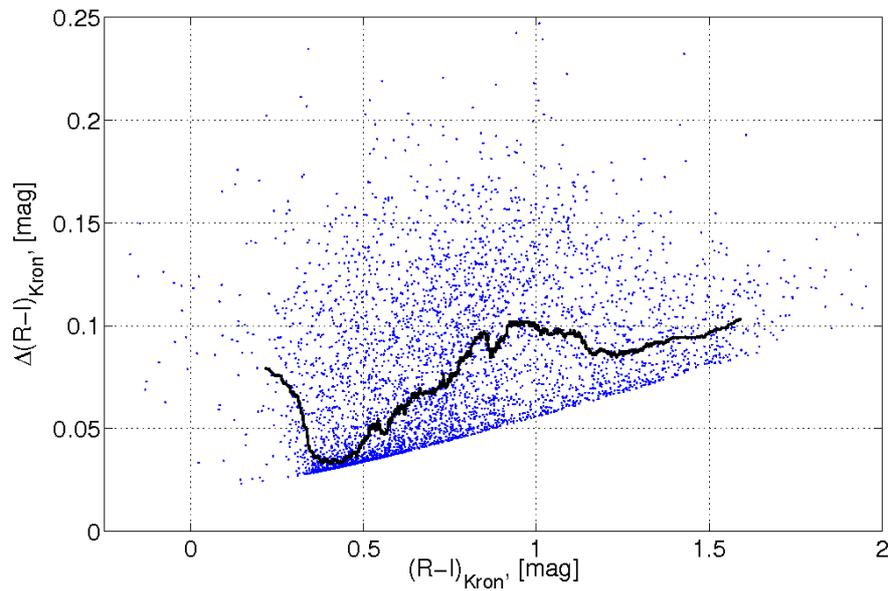}
\end{center}
\caption{(R - I) colour errors as a function of (R - I) colour for $\sim$4,000 objects imaged in the same field. The black
line shows a running median of errors in 200-object bins.  
   \label{Fig:RIdRI}}
\end{figure*}

Excluding all the sources with R $>$ 20.6 mag and I $>$ 19.6, we expect that the RMS error of the (R - I) colour should be 
smaller than $(0.15)\times\sqrt{2}\cong0.2$ mag. Fig. \ref{Fig:RIdRI} shows the (R - I) colour errors as a function of 
(R - I) colour for $\sim$4,000 objects brighter than R = 20.6 and I = 19.6 mag, extracted from the same 300-sec combined
exposure as in Figs. \ref{Fig:RdR} and \ref{Fig:IdI}. The median of the colour errors is $\cong0.1$ mag for colour 
indices $-0.25\leq(R-I)\leq2.2$.
We therefore adopt R$_{lim} = 20.6$ mag and I$_{lim} = 19.6$ mag as the limiting magnitudes of the NCCS.

\section{Point/Extended Source Separation}

In this section we describe an empirical point/extended source separation (PES) procedure for the NCCS images. Below we refer
to ``extended'' sources implying that they are very likely to be non-stellar, probably galaxies. The procedure 
is based on the SDSS star/galaxy classification.

To compare our results with the SDSS and to perform PES we obtained three 
300-sec images in the R and in I filters of a field covered by the SDSS and centered on J2000.0 
($\alpha$, $\delta$) = (16$^h$, 40$^{\circ}$) with exactly the same setup as used for the NCCS. 
The images were debiased, FF normalized and median-combined. After deriving the astrometric solution for 
the resulting images, the objects were extracted using the SE and were matched with the SDSS objects.

\begin{figure*}[ht!]
\begin{center}
\begin{tabular}{cccc}
(a) & \includegraphics[angle=0,width=70mm]{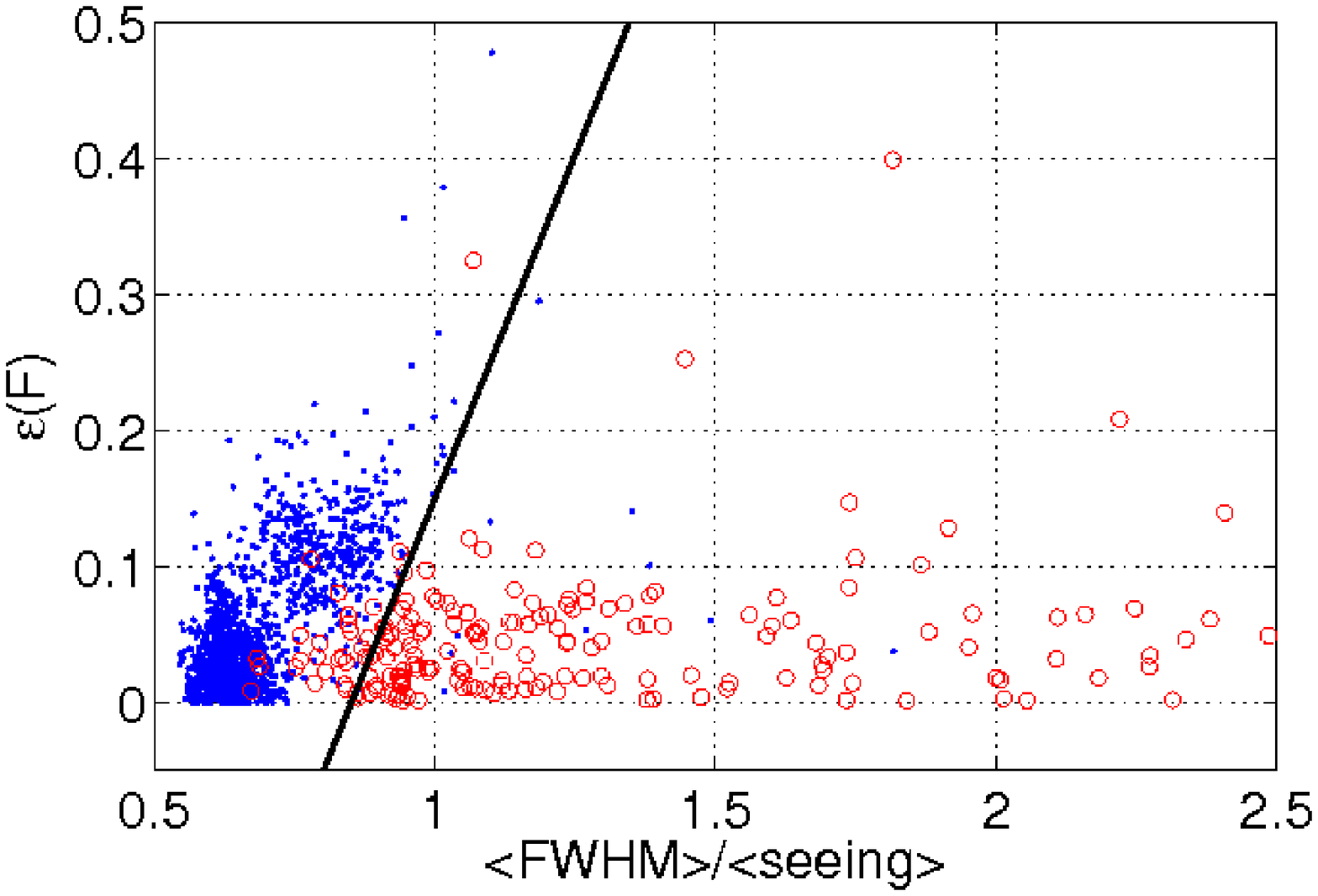} &
(b) & \includegraphics[angle=0,width=70mm]{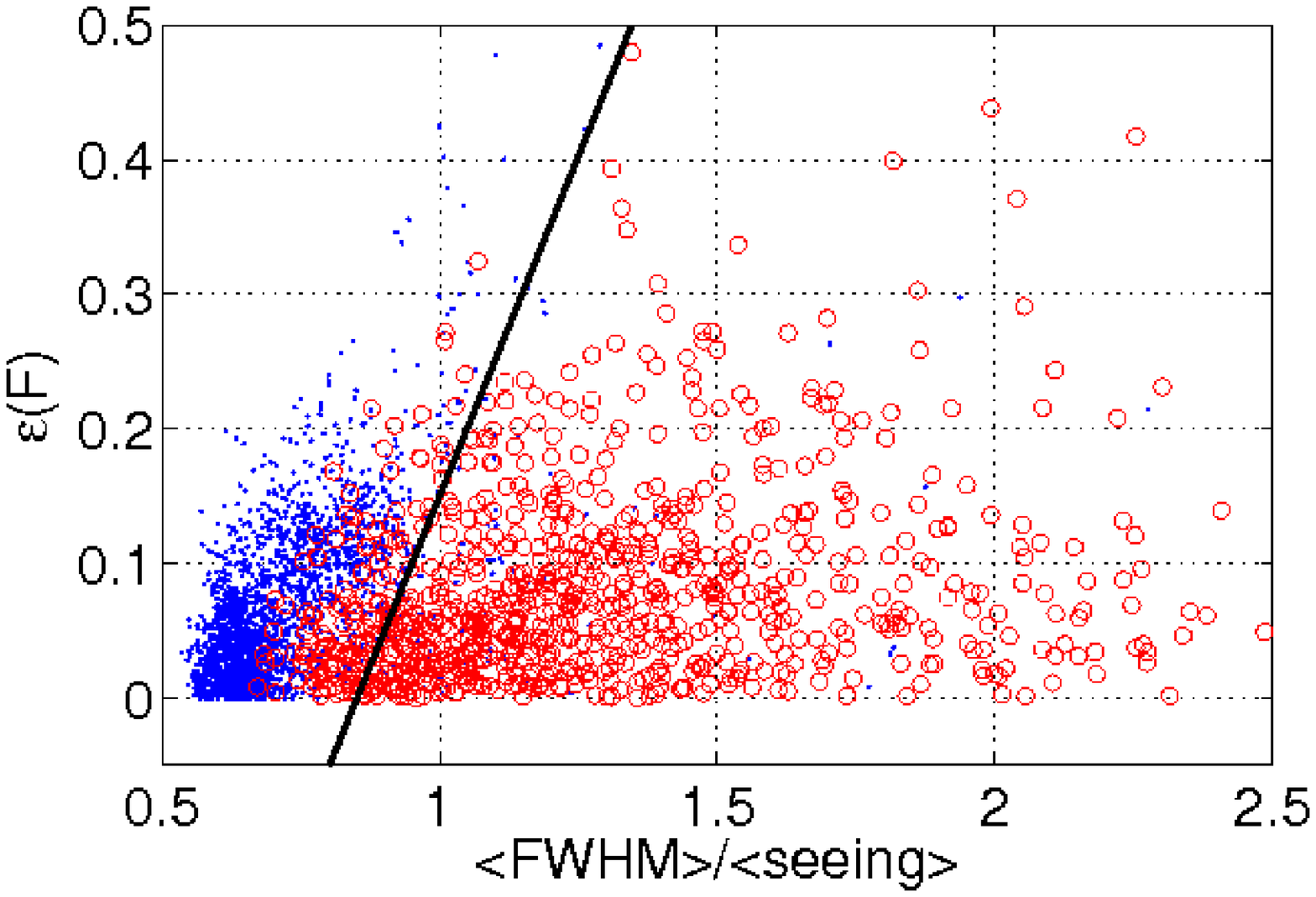} \\
(c) & \includegraphics[angle=0,width=70mm]{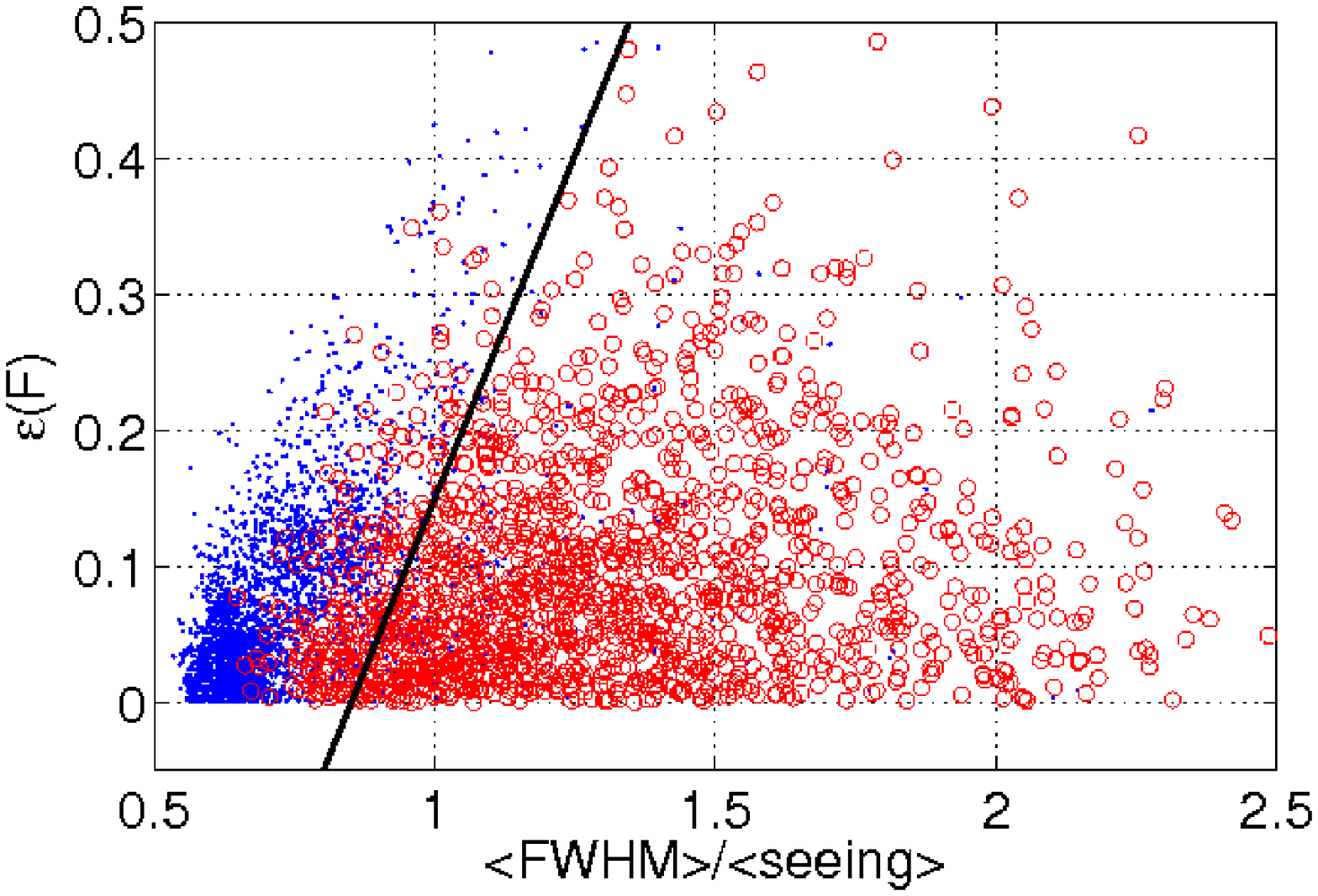} & &
\end{tabular}
\end{center}
\caption{Relative FWHM difference $\epsilon$(FWHM) as a function of (mean FWHM)/(mean seeing). Points show objects defined 
by SDSS as point 
sources and circles show objects defined as extended by SDSS. The solid black line represents an empirical PES solution 
defined by equation (\ref{Eq:PES}).
Panel (a) shows objects brighter than R$=18.6$ mag and I$=17.8$ mag. Panel (b) - objects brighter than R$=20.0$ mag and I$=19.1$ mag.
Panel (c) - objects brighter than R$=20.6$ mag and I$=19.6$ mag.
   \label{Fig:PES}}
\end{figure*}

Our PES procedure is based on the size and shape of the sources. 
We defined the relative FWHM difference in R and I bands as $\epsilon$(FWHM)$=\frac{|FWHM_R-FWHM_I|}{FWHM_R+FWHM_I}$. We also defined 
the mean FWHM in R and I bands as $<FWHM>=\frac{FWHM_R+FWHM_I}{2}$. Fig. \ref{Fig:PES} shows plots of $\epsilon$(FWHM) vs. mean FWHM normalized by the mean seeing, for different magnitude limits. Note that
the objects can be separated with reasonable accuracy using a single separation line. Different separation lines were examined using two
criteria: the separation needs to match the SDSS PES as good as possible, and the mutual contaminations by point and extended sources need
to be similar, i.e. with no bias to one of the classes. The first criterion implies that the number of the point/extended sources
classified incorrectly by our routine relative to 
the SDSS should be as small as possible. The second criterion implies that the number of the point sources defined incorrectly 
by our routine should be similar to the number of the extended sources defined incorrectly.

The following PES solution fits best these two criteria:
\begin{equation}
\begin{array}{rcll} 
 \epsilon(FWHM) & > & 1.01\frac{<FWHM>}{<seeing>} - 0.86& \rightarrow point~ source \\
 \epsilon(FWHM) & < & 1.01\frac{<FWHM>}{<seeing>} - 0.86& \rightarrow extended~ source
\end{array}
\label{Eq:PES}
\end{equation}

Table \ref{Tab:PES} shows the PES accuracy relative to SDSS obtained using equation (\ref{Eq:PES}) for different magnitude limits. 
The PES classification accuracy corresponding to the NCCS limiting magnitudes R$_{lim} = 20.6$ mag and I$_{lim} = 19.6$ mag,
as defined in Section 6, is $\sim$91\%.
Since the separation requires that both the R and I band data would be available for the same object,
the PES is performed following stage two of the photometric procedure. 
The final catalogue includes the PES flag defined for each object, which is 0 for a point source and 1 for an extended one. 

\begin{table}[htbp]
\caption{PES Accuracy Relative to the SDSS Corresponding to the Different R and I Limits:}
\begin{center}
\begin{tabular}{c|c|c}
\hline
\multicolumn{ 2}{c|}{Magnitude Limits} &  \\ \hline
~~R~~ & I & PES Accuracy \\ \hline
~~~18.6~~~ & 17.8 & 97.1\% \\ 
~~~20.0~~~ & 19.1 & 93.0\% \\ 
~~~20.6~~~ & 19.6 & 90.8\% \\ \hline
\end{tabular}
\end{center}
\label{Tab:PES}
\end{table}

\section{Comparison with the SDSS}

Here we compare the results of our observations, reductions and extraction procedures with data provided by the SDSS.
Using the same SDSS field images from June 12, 2009 we compare the NCCS photometry and astrometry with the SDSS. \cite{JOR06} 
derived empirical 'global' colour transformations between SDSS photometry and Johnson-Cousins photometric system for $\sim$4,000
standard stars. To obtain \textit{r} and \textit{i} band magnitudes of the NCCS objects, we use the following equations from 
\cite{JOR06}:
\begin{equation}
\begin{array}{rcl} 
i-I & = & (0.247 \pm0.003)*(R-I)  + (0.329 \pm 0.002)\\
r-i & = & (1.007 \pm0.005)*(R-I)  - (0.236 \pm 0.003) 
\end{array}
\label{Eq:Jordi}
\end{equation}

\begin{figure*}[ht!]
\begin{center}
\begin{tabular}{cccc}
(a) & \includegraphics[angle=0,width=70mm]{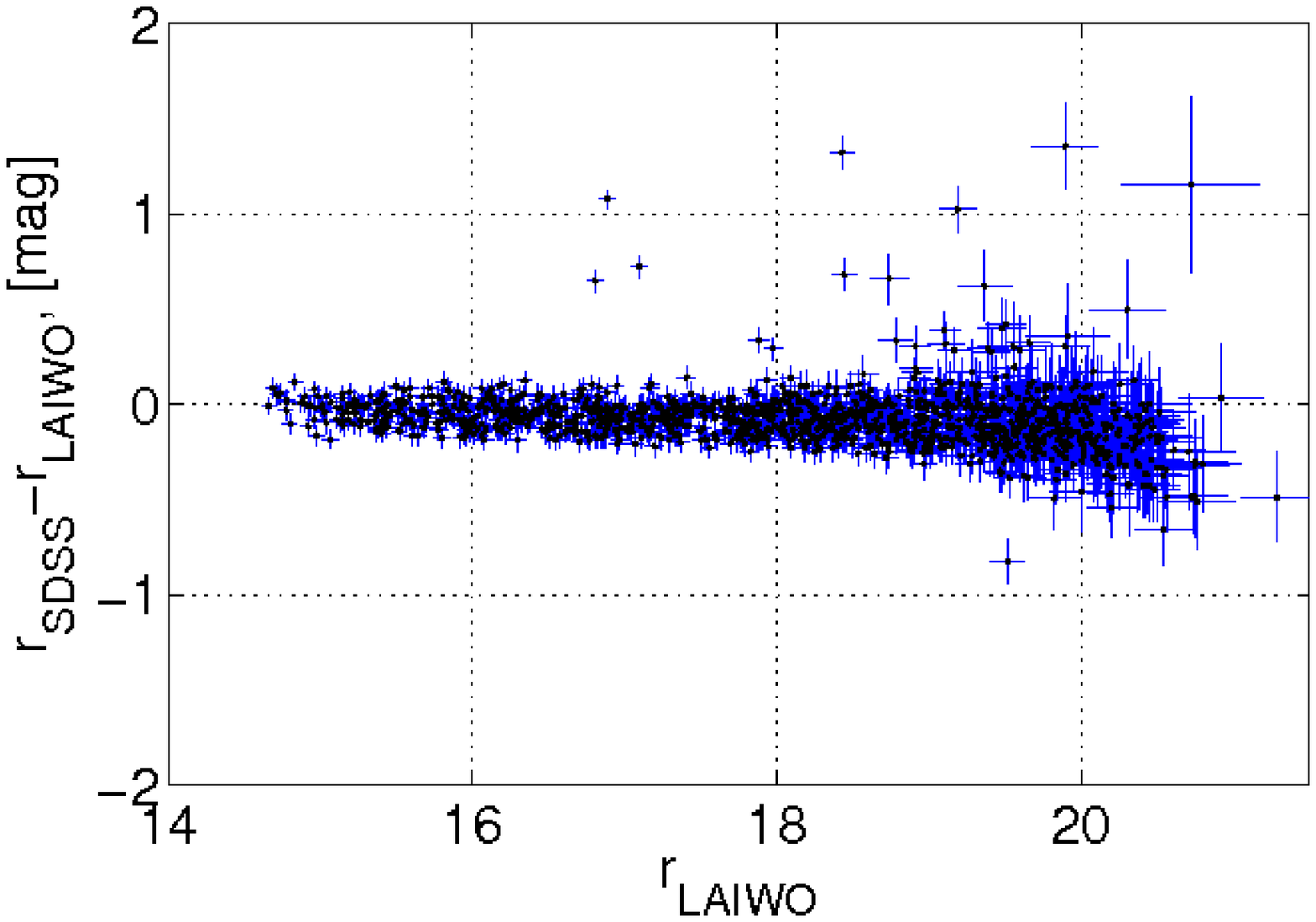} &
(b) & \includegraphics[angle=0,width=70mm]{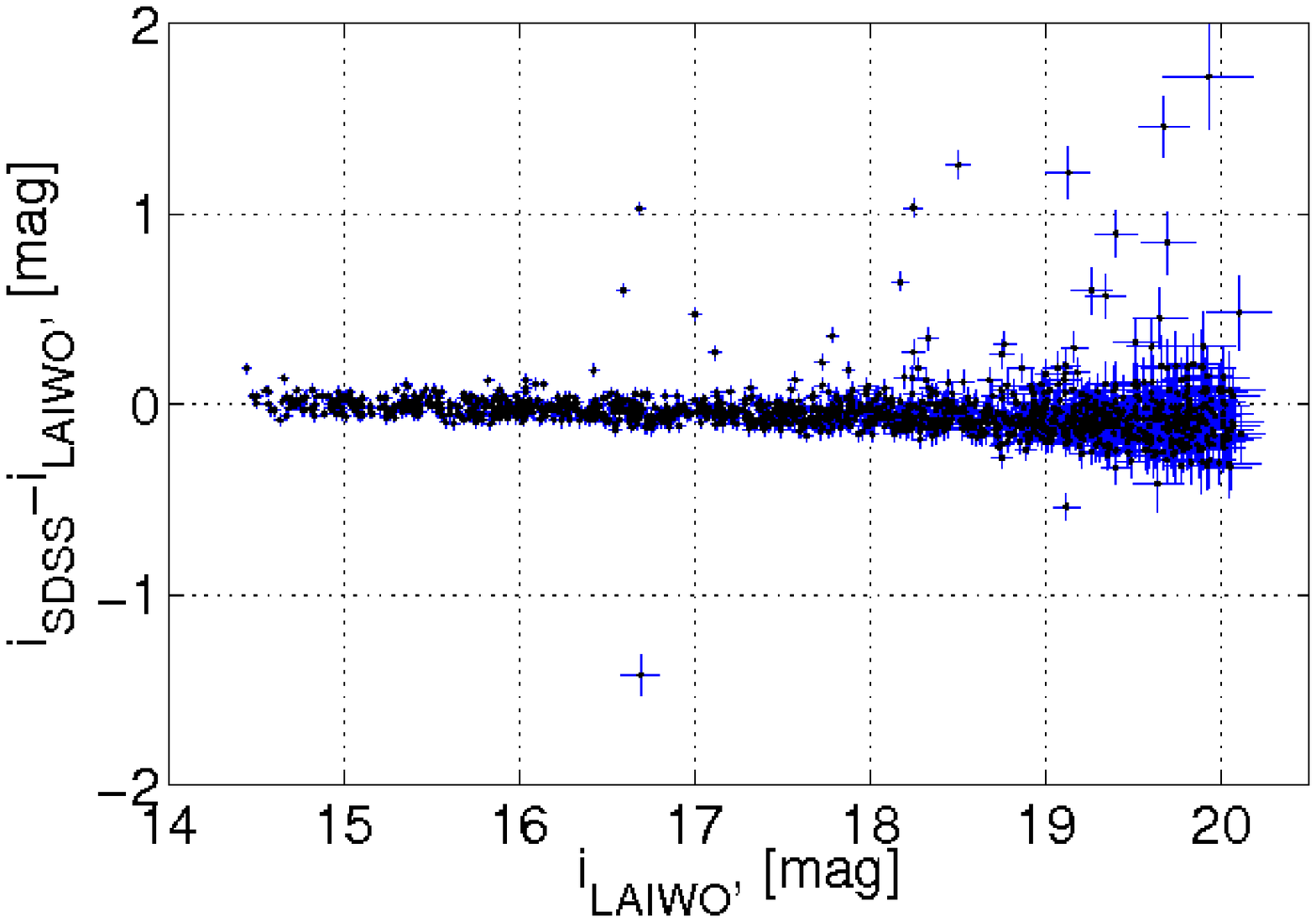} \\
(c) & \includegraphics[angle=0,width=70mm]{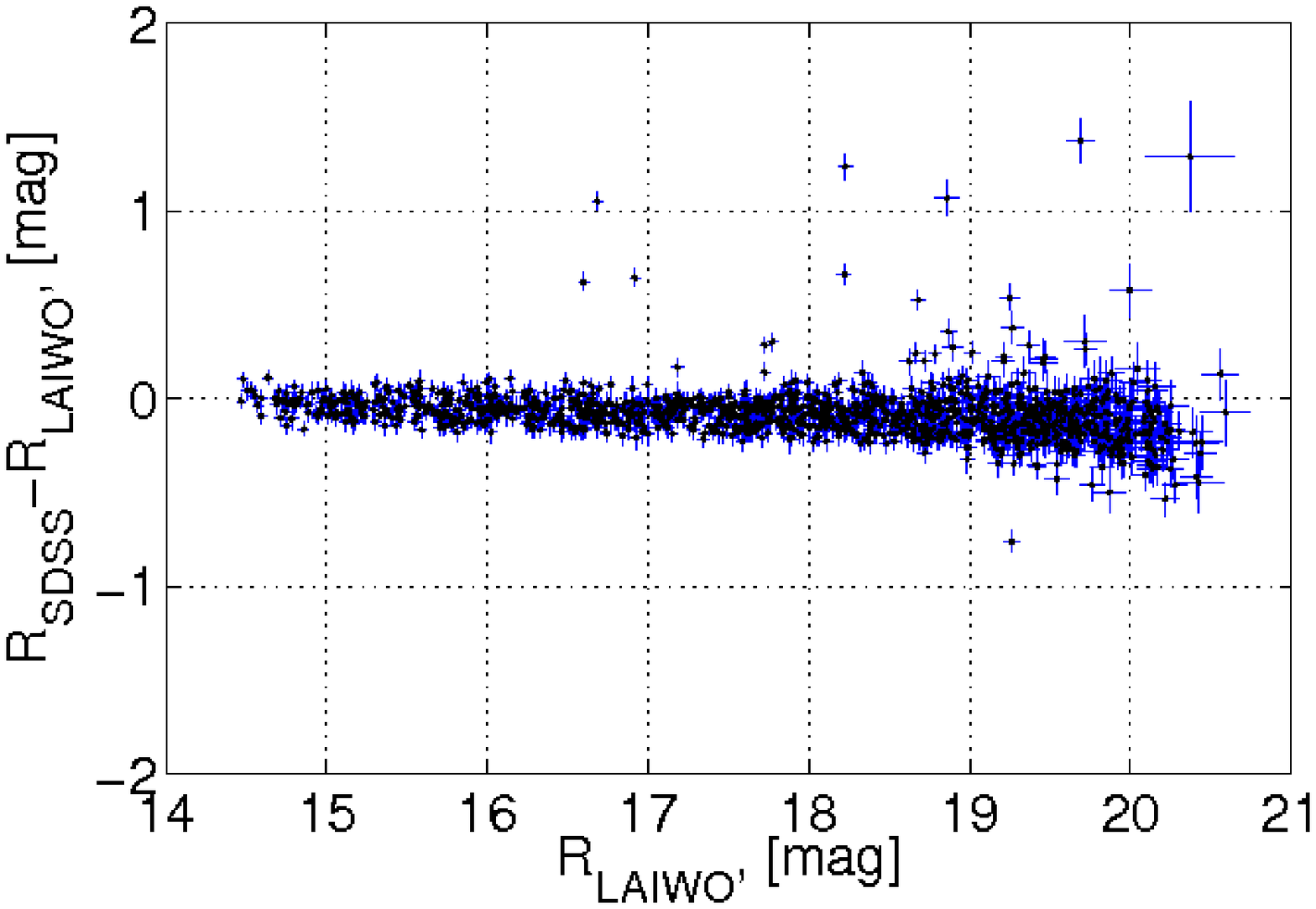} & 
(d) & \includegraphics[angle=0,width=70mm]{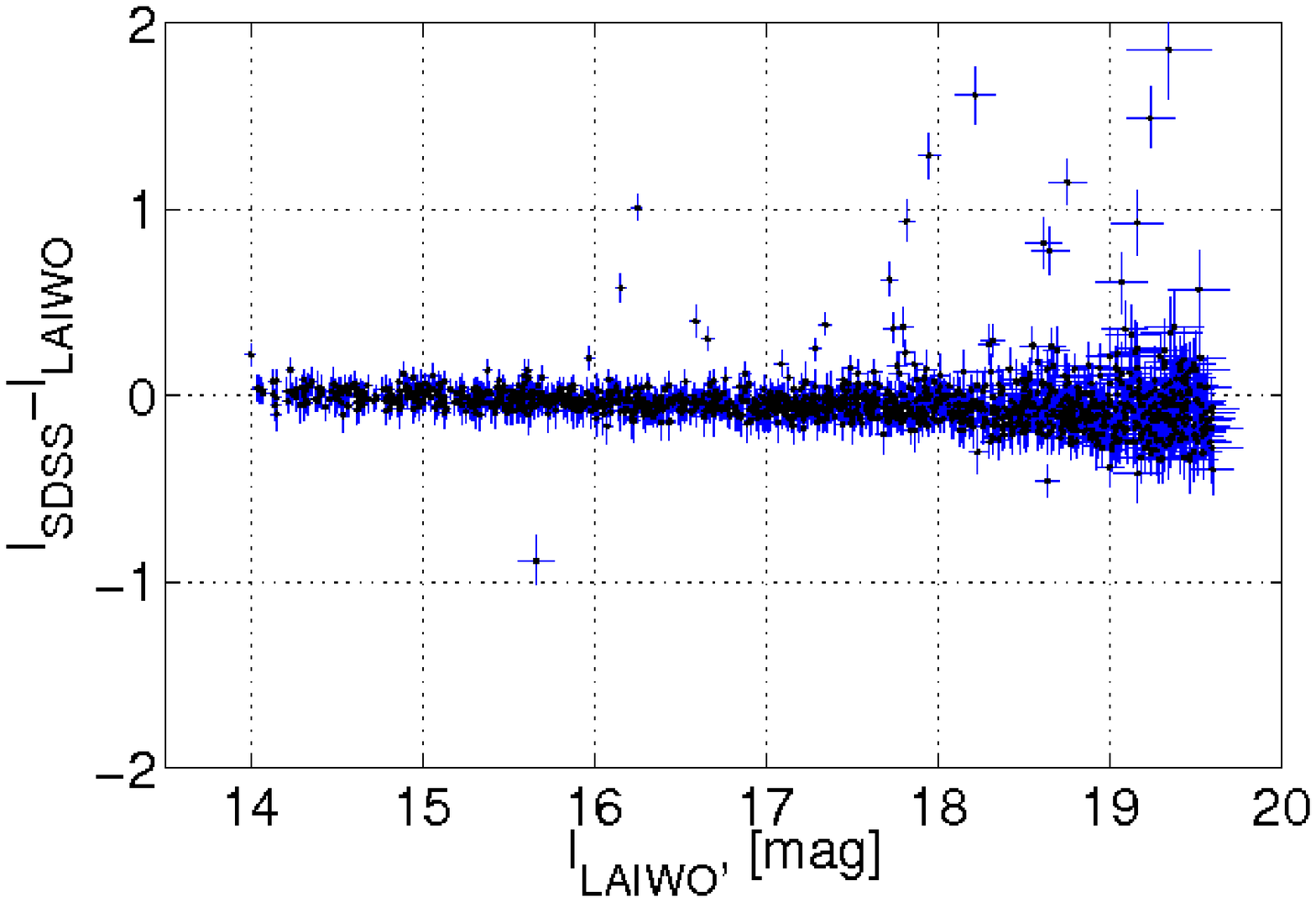}
\end{tabular}
\end{center}
\caption{The NCCS photometry compared to the SDSS photometry. Panels (a) and (b) show \textit{r} and \textit{i} 
magnitudes comparison derived using the \cite{JOR06} transformation. Panels (c) and (d) show R and I magnitudes comparison 
derived using the \cite{CHO08} transformation.
   \label{Fig:NCCSSDSSphot}}
\end{figure*}

Panels (a) and (b) in Fig. \ref{Fig:NCCSSDSSphot} show the comparison between the NCCS photometry, transformed to
\textit{r} and \textit{i} magnitudes using equations (\ref{Eq:Jordi}), and the SDSS photometry for the objects 
brighter than the NCCS limiting magnitudes R$_{lim} = 20.6$ mag and I$_{lim} = 19.6$ mag,
as defined in Section 6. The NCCS and SDSS photometric results correlate 
well; Table \ref{Tab:RMSphot} shows the RMS deviations of the NCCS photometry relative to the SDSS photometry
for different magnitude limits. Note that the RMS deviations in both bands become smaller for brighter magnitudes.

The RMS deviation values in Table \ref{Tab:RMSphot} are greater than the RMS errors for the NCCS limiting magnitudes
derived in Section 6, but are still comparable. The difference is explained by the transformation uncertainty 
in equations (\ref{Eq:Jordi}). The transformation uncertainty can be defined as the dispersion of the data points used to derive
the transformation, which we estimate as $\sigma\sim 0.1$ mag. Therefore, the RMS deviation in the \textit{i} band for the stars 
brighter than I$ = 19.6$ mag is $\sqrt{0.1^2+0.15^2}\cong0.18$ mag, which is consistent with the value for I = 19.6 shown in 
Table \ref{Tab:RMSphot}. The first term under the square root sign is produced by the RMS deviation of the transformation, 
while the second term is produced 
by the RMS magnitude error of the stars with I = 19.6. The \textit{r} band transformation in \cite{JOR06} is determined using
the \textit{i} band transformation, thus the transformation uncertainty $\sigma\sim 0.1$ needs to be accounted for twice. 
This results in a greater dispersion 
of the data points in the \textit{r} band plot in Fig. \ref{Fig:NCCSSDSSphot} compared to that of the data points in the 
\textit{i} band plot. The RMS deviation in the \textit{r} band for stars 
brighter than R = 20.6 mag is $\sqrt{0.1^2+0.1^2+0.15^2}\cong0.21$ mag, which is consistent with the value for R = 20.6
shown in Table \ref{Tab:RMSphot}. The RMS deviation estimations for the other magnitude limits produce the values similar to 
those shown in Table \ref{Tab:RMSphot}.

The shortcomings of \cite{JOR06} transformations were pointed out by \cite{CHO08}, who proposed their own transformations 
from the SDSS \textit{ugriz} magnitudes to the Johnson-Cousins UBVRI magnitudes. We performed an additional check on the quality
of our 
photometry calibration using the following equations from \citeauthor{CHO08}:
\begin{equation}
\begin{array}{rcl} 
R & = & r - (0.272 \pm0.092)*(r-i)  - (0.159 \pm 0.022)\\
I & = & i - (0.337 \pm0.191)*(r-i)  - (0.370 \pm 0.041) 
\end{array}
\label{Eq:Chonis}
\end{equation}

The comparison between the NCCS photometry and the SDSS photometry, transformed to the R and I magnitudes
using equations (\ref{Eq:Chonis}), is shown in panels (c) and (d) of Fig. \ref{Fig:NCCSSDSSphot}. The RMS deviations 
of the NCCS photometry relative to the SDSS photometry for different magnitude limits are presented in Table \ref{Tab:RMSphot}.
The RMS deviation values in Table \ref{Tab:RMSphot} are greater than the RMS errors for the NCCS limiting magnitudes
derived in Section 6, but are still comparable and become smaller for brighter magnitudes. The difference is explained by 
the uncertainty of the transformation in equations (\ref{Eq:Chonis}) as described above. Note that the data dispersion 
in panels (c) and (d) of Fig. \ref{Fig:NCCSSDSSphot} is similar, since the transformation equations (\ref{Eq:Chonis}) 
for R and I bands are independent. Note also that there seems to be a turndown in the plots of Fig. \ref{Fig:NCCSSDSSphot}
for stars with I $\gtrsim 19$, R $\gtrsim 19$ mag and $i \gtrsim 19.5$, $r \gtrsim 19.5$ mag, which can be attributed by a
Malmquist bias, as explained by \cite{CHO08}.


\begin{table}[htbp]
\caption{RMS Deviations of the NCCS Photometry Relative to the SDSS Photometry Corresponding to the Different R and I Limits:}
\begin{center}
\begin{tabular}{c|c||c|c|c|c}
\hline
\multicolumn{ 2}{c||}{Magnitude Limits} & \multicolumn{ 4}{c}{RMS Deviation (SDSS - LAIWO)} \\ \hline
\multicolumn{ 2}{c||}{ } & \multicolumn{ 2}{c|}{Jordi et al.} & \multicolumn{ 2}{c}{Chonis \& Gaskell} \\ \hline
~~R~~ & I & $\sigma$(\textit{r}) & $\sigma$(\textit{i}) & $\sigma$(R) & $\sigma$(I) \\ \hline
~~~18.6~~~ & 17.8 & ~~~0.17~~~ & ~~~0.10~~~ & ~~~0.16~~~ & 0.09 \\ 
~~~20.0~~~ & 19.1 & ~~~0.20~~~ & ~~~0.14~~~ & ~~~0.18~~~ & 0.14 \\ 
~~~20.6~~~ & 19.6 & ~~~0.21~~~ & ~~~0.17~~~ & ~~~0.19~~~ & 0.17 \\ \hline
\end{tabular}
\end{center}
\label{Tab:RMSphot}
\end{table}

\begin{figure*}[ht!]
\begin{center}
\begin{tabular}{cccc}
(a) & \includegraphics[angle=0,width=70mm]{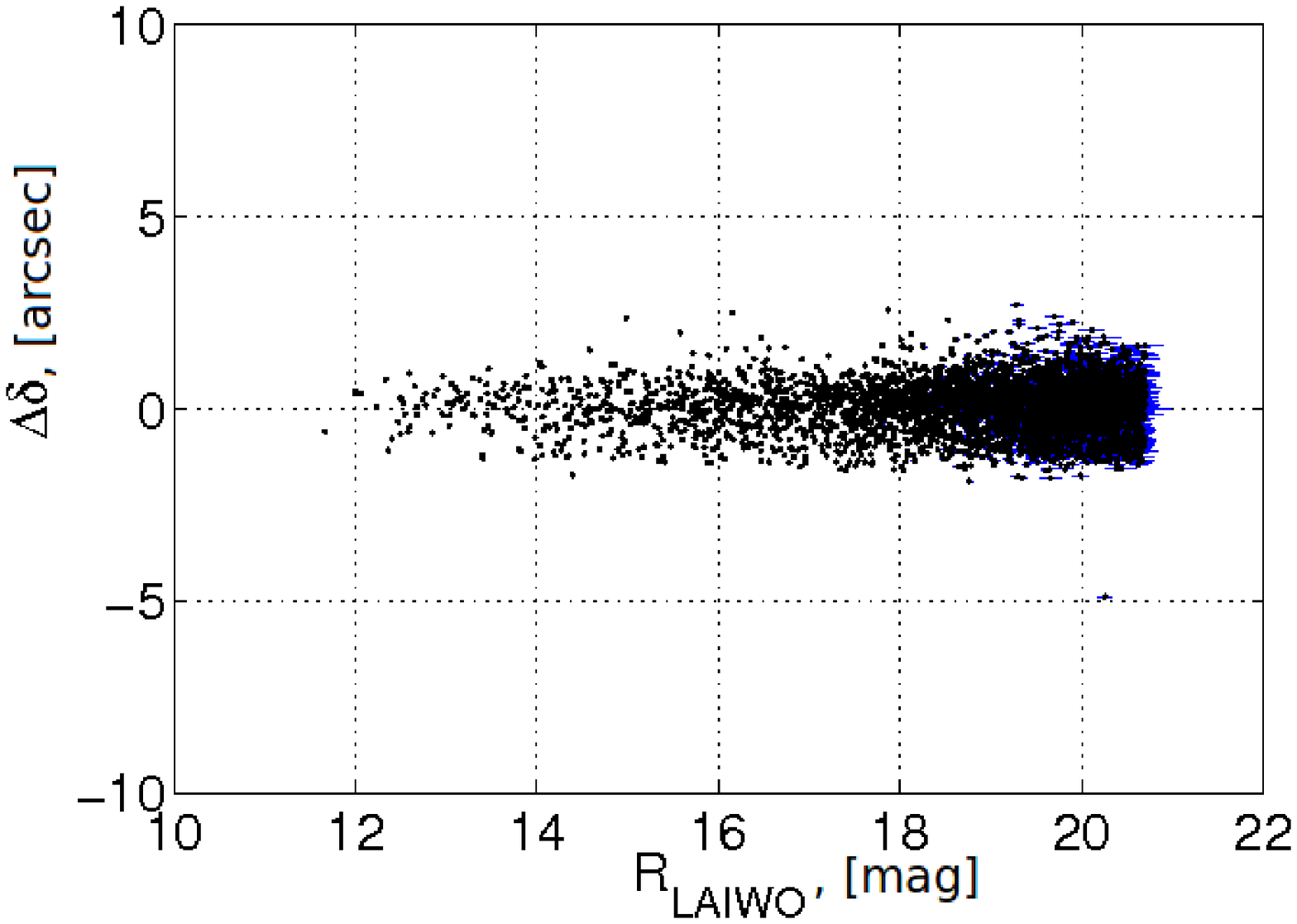} &
(b) & \includegraphics[angle=0,width=70mm]{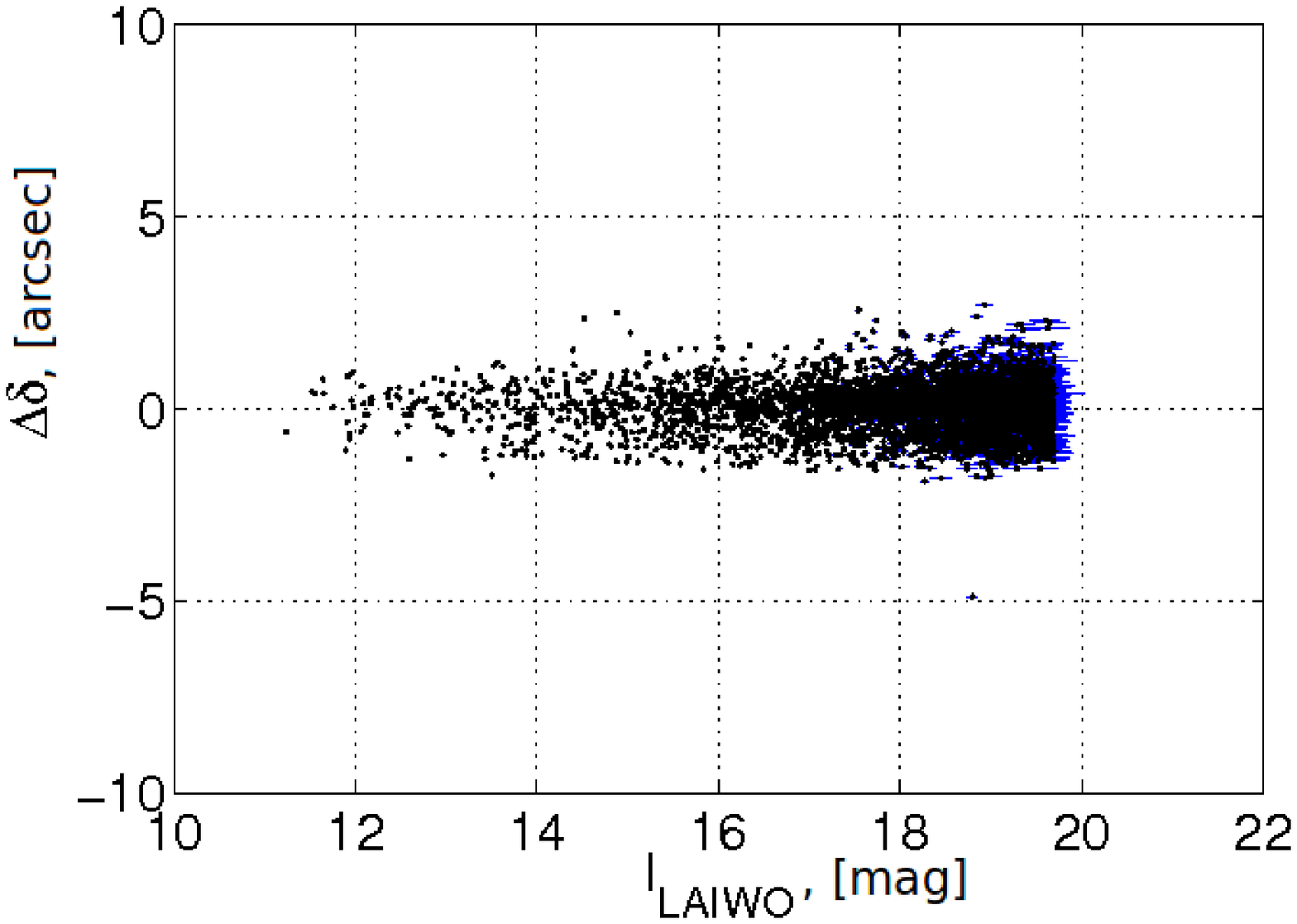} \\
(c) & \includegraphics[angle=0,width=70mm]{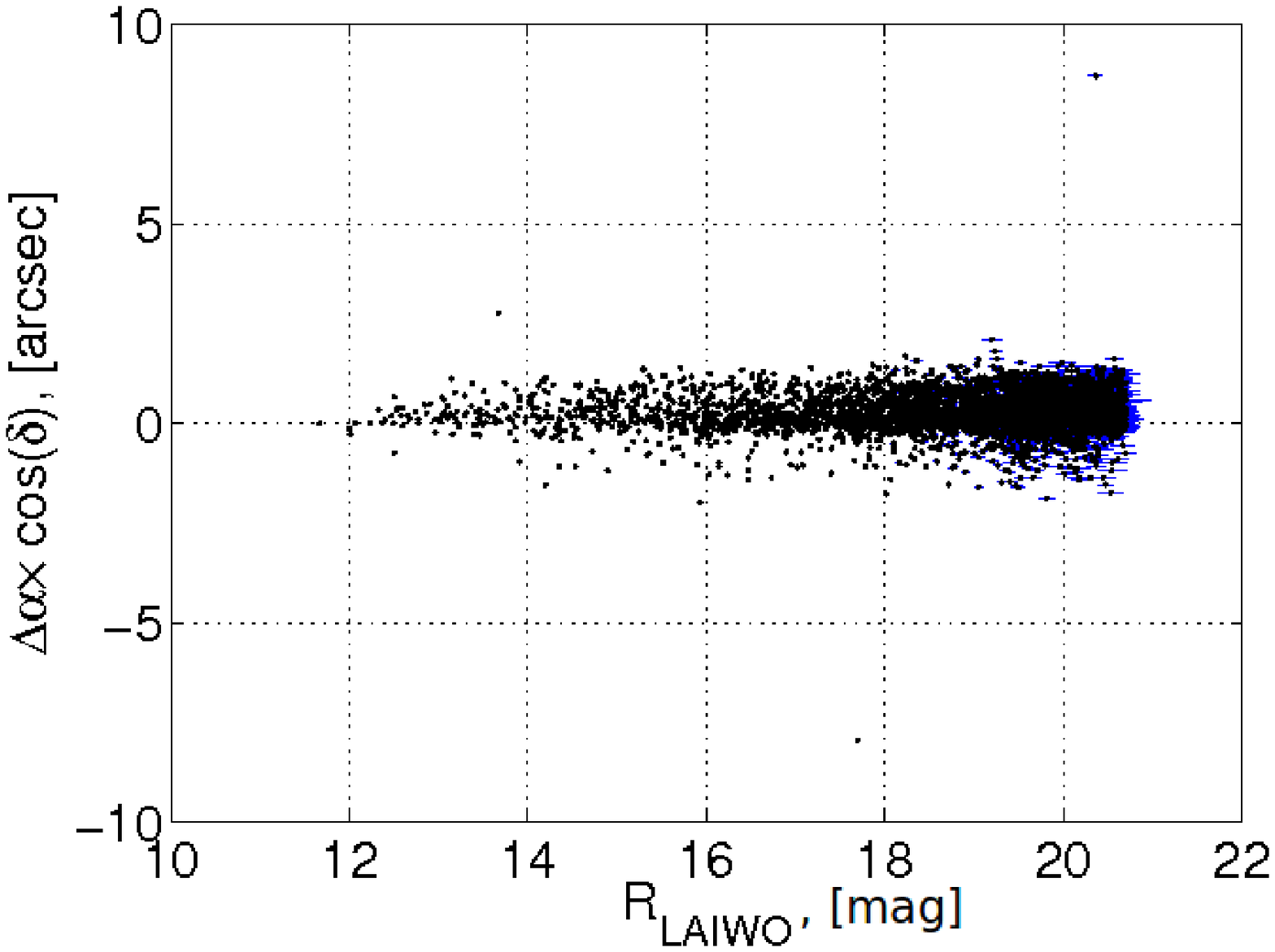} & 
(d) & \includegraphics[angle=0,width=70mm]{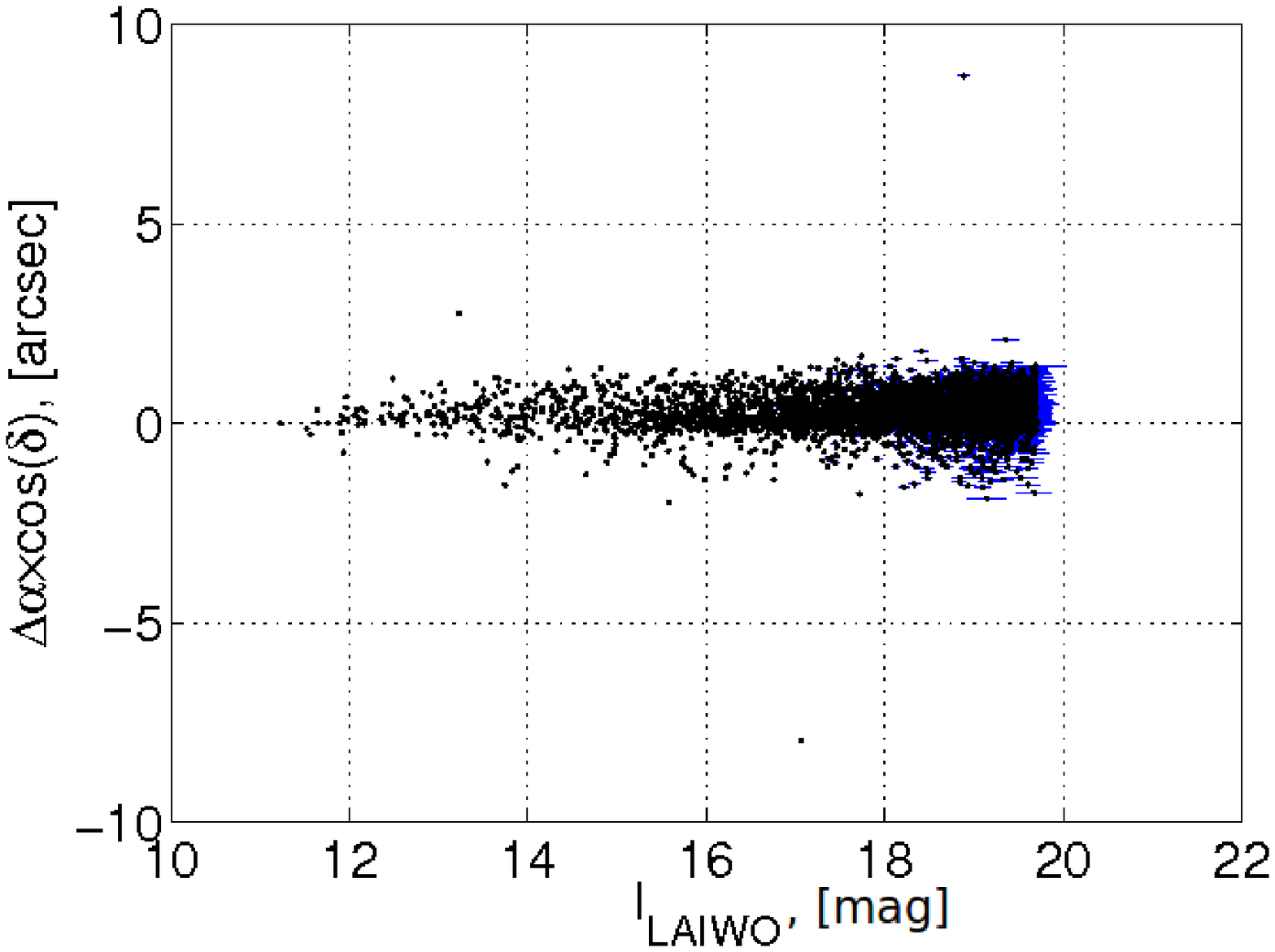}
\end{tabular}
\end{center}
\caption{The NCCS astrometry compared to the SDSS astrometry. Panels (a) and (b) show the $\delta$ deviations. 
Panels (b) and (c) show $\alpha\times$cos($\delta$) deviations.
   \label{Fig:NCCSSDSSastro}}
\end{figure*}

Fig. \ref{Fig:NCCSSDSSastro} shows the comparison between the NCCS astrometry and the SDSS astrometry for the objects 
brighter than the NCCS limiting magnitudes R$_{lim} = 20.6$ mag and I$_{lim} = 19.6$ mag. The following RMS deviations 
were obtained: $\sigma(\Delta\delta) \cong 1.12$ arcsec 
and $\sigma(\Delta\alpha\times\cos\delta) \cong 0.83$ arcsec. 
These RMS deviation values are smaller than the NCCS astrometric solution RMS errors derived in Section 5.

\begin{figure*}[ht!]
\begin{center}
\begin{tabular}{cccc}
(a) & \includegraphics[angle=0,width=70mm]{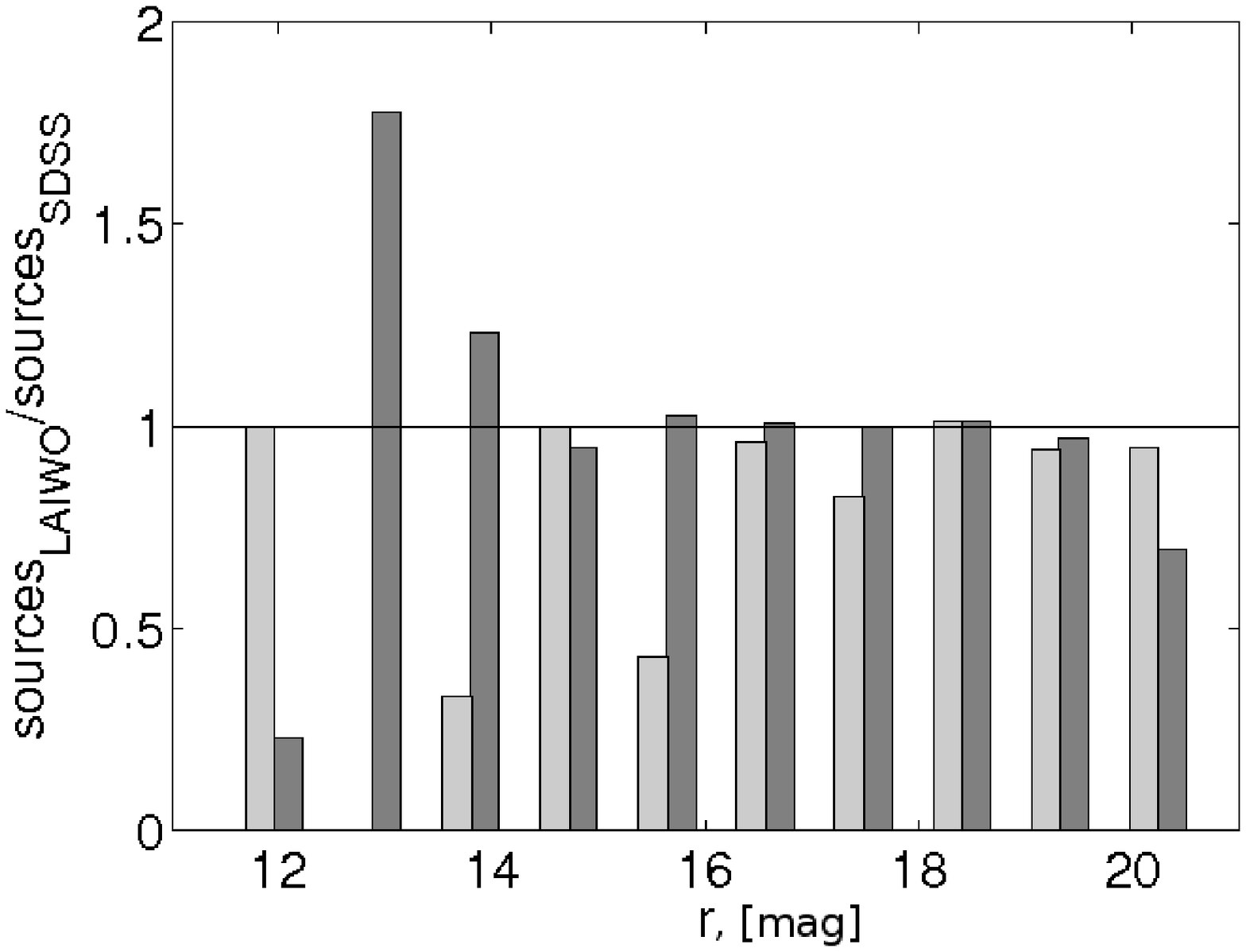} &
(b) & \includegraphics[angle=0,width=70mm]{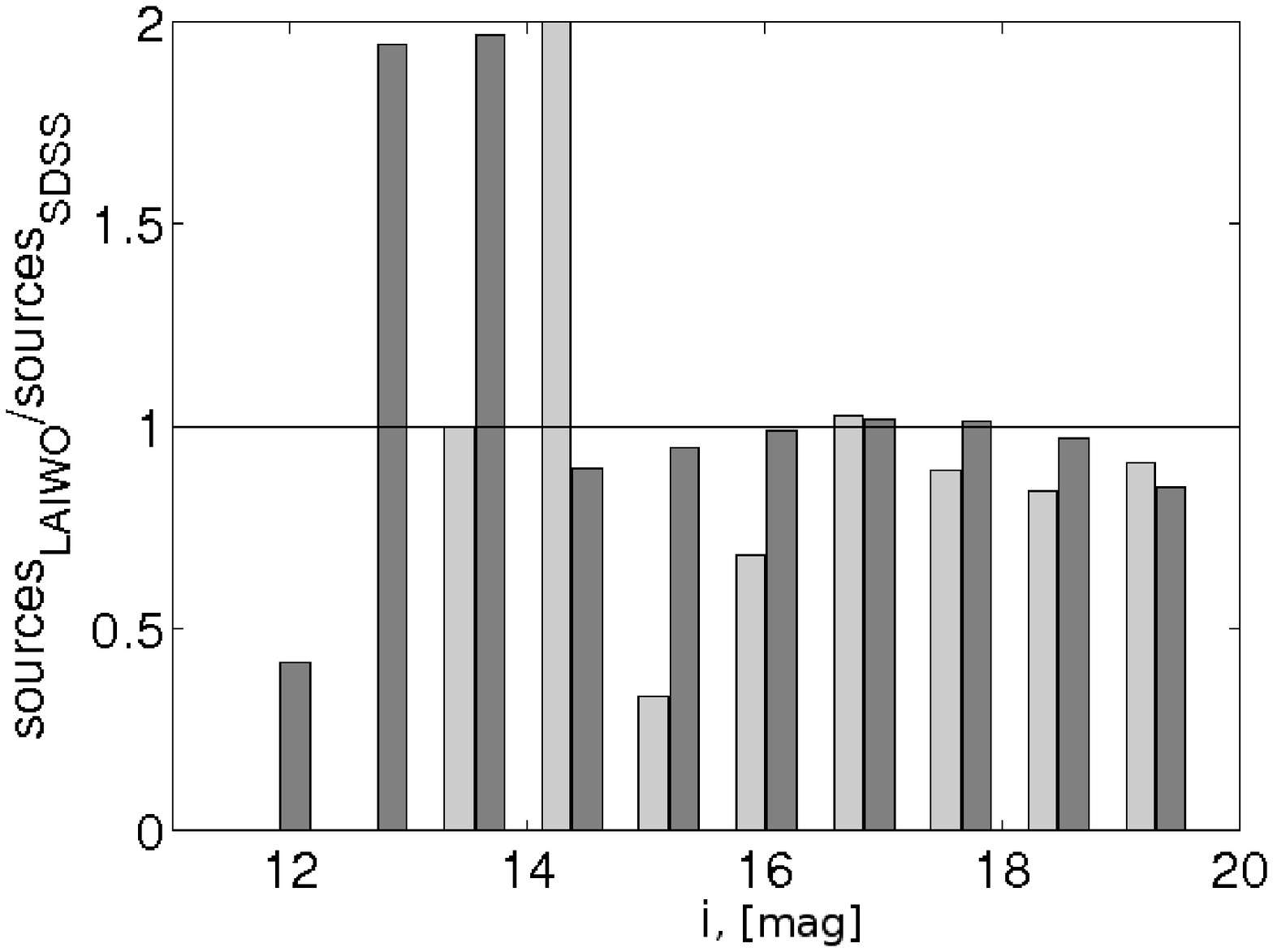} 
\end{tabular}
\end{center}
\caption{Number of detected souces ratio as a function of SDSS magnitude. 
Dark grey bars represent point sources. Light grey bars represent extended sources. Panel (a) is for
the $r$ magnitude. Panel (b) is for the $i$ magnitude.
   \label{Fig:sources}}
\end{figure*}

Fig. \ref{Fig:sources} shows the ratio between the number of sources detected by NCCS and the number of sources detected by SDSS in 
the same field, as a function of the \textit{r} and \textit{i} magnitudes. NCCS detects more sources brighter than 
$r=14.5$ and $i=14.5$ mag than the SDSS does. 
This is probably due to the saturation limit of the SDSS, which is $r=14.5$ and $i=14.5$ mag \citep{CHO08}. NCCS detects $\sim$100\% 
of the point sources and $\sim$90\% of the extended sources fainter than $r=14.5$ and $i=14.5$ mag (but brighter 
than the NCCS limits) detected by the SDSS in both $r$ and $i$ bands. 
Small deviations of the ratio from 100\% are probably due to uncertainties of the photometric 
transformation in equation (\ref{Eq:Jordi}). Note also that the detection ratio drops for the sources fainter 
than $r=20$ and $i=19$ mag, which are close to the NCCS limiting R and I magnitudes defined in Section 6 and can be attributed 
by a Malmquist bias as explained by \cite{CHO08}..

\section{Sky Coverage and Completeness}

\begin{figure*}[ht!]
\begin{center}
\includegraphics[angle=0,width=120mm]{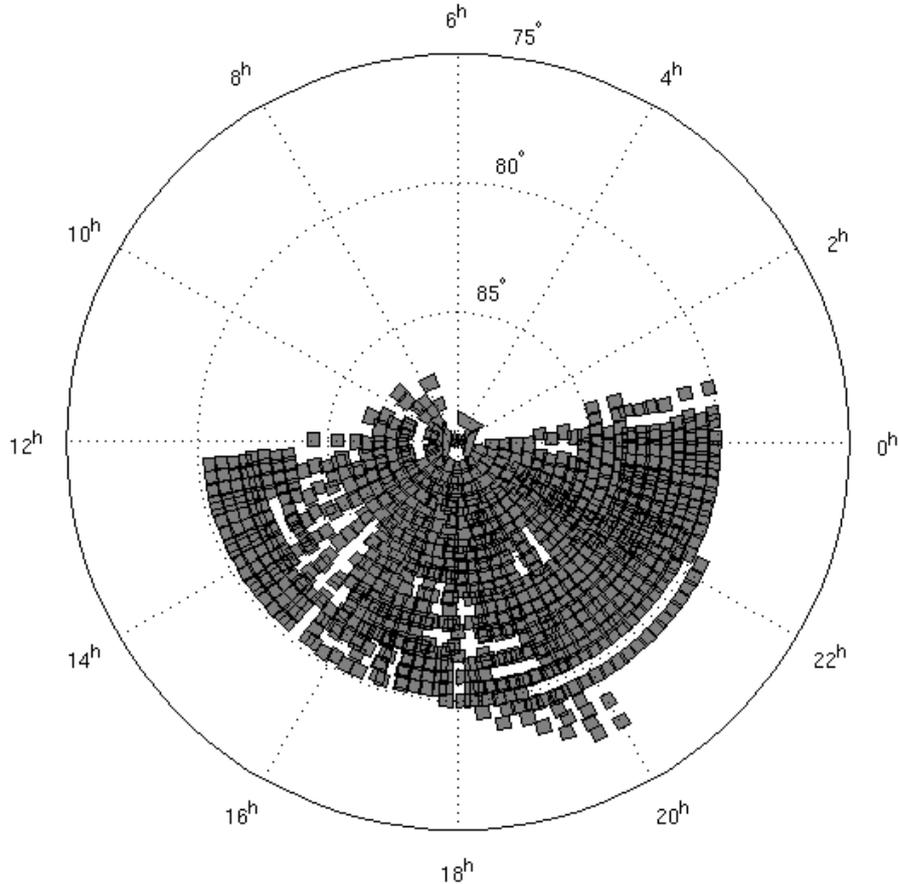}
\end{center}
\caption{Sky coverage from February 13, 2009 till September 13,2009.
   \label{Fig:coverage}}
\end{figure*}

Since February 13, 2009 we imaged 223 sky fields three times in $R$ and $I$ filters in the NCC region for a total 
sky coverage of $\sim$130 square degrees. Fig. \ref{Fig:coverage} shows the sky coverage map in polar projection with each
square representing the footprint of one of LAIWO science CCDs. The survey results will be described in a future paper.

\begin{figure*}[ht!]
\begin{center}
\begin{tabular}{cccc}
(a) & \includegraphics[angle=0,width=70mm]{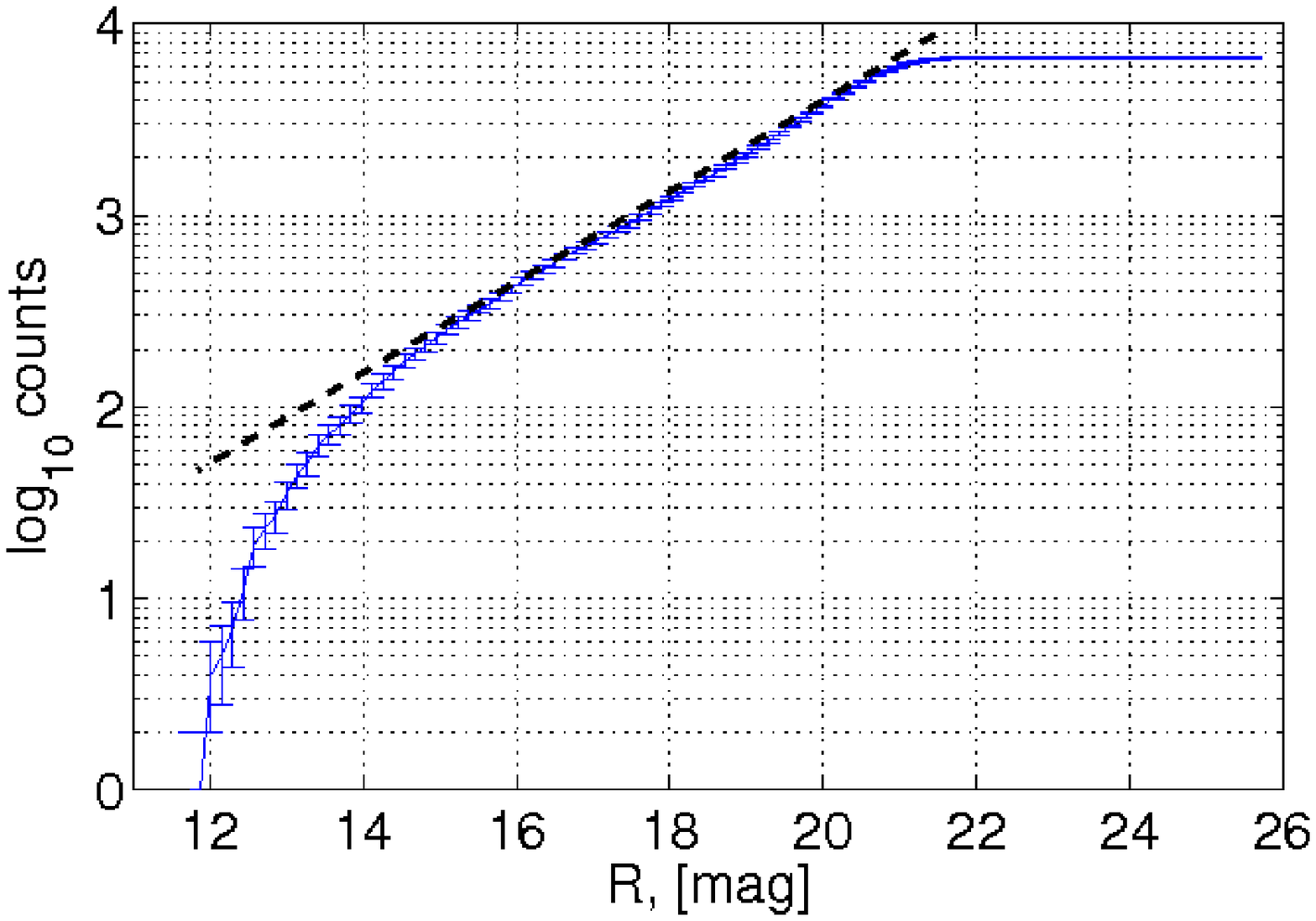} &
(b) & \includegraphics[angle=0,width=70mm]{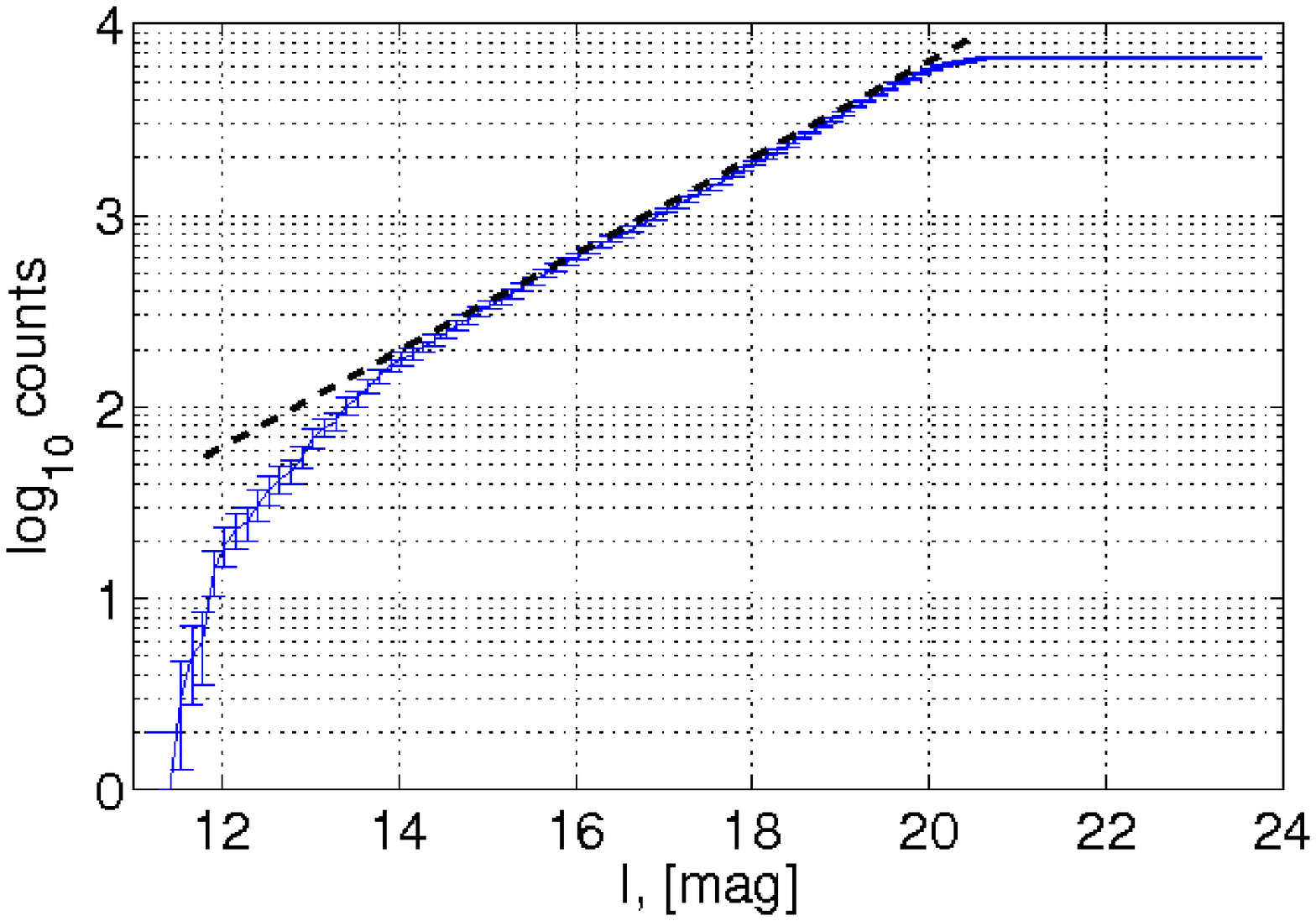} 
\end{tabular}
\end{center}
\caption{Star count cumulative distribution as a function of magnitude. The black dashed line shows an exponential model
of a complete catalogue. The left panel is for
the R magnitudes and the right panel is for the I magnitudes.
   \label{Fig:complete}}
\end{figure*}

We use images from June 12, 2009 to estimate the catalogue completeness. Fig. \ref{Fig:complete} shows a comparison of the star
count cumulative distribution with an exponential model expected for a complete catalogue. Note that no sources are missing for 
R $\leq20.7$ mag and I $\leq19.9$ mag. These completeness limits are fainter than the limiting R and I magnitudes as defined
in Section 6. Therefore, we estimate that the NCCS catalogue will be complete to the limiting magnitudes as defined in Section 6.

\section{Multiple Detections and Variability Treatment}

\begin{figure*}[ht!]
\begin{center}
\begin{tabular}{cccc}
(a) & \includegraphics[angle=0,width=73mm]{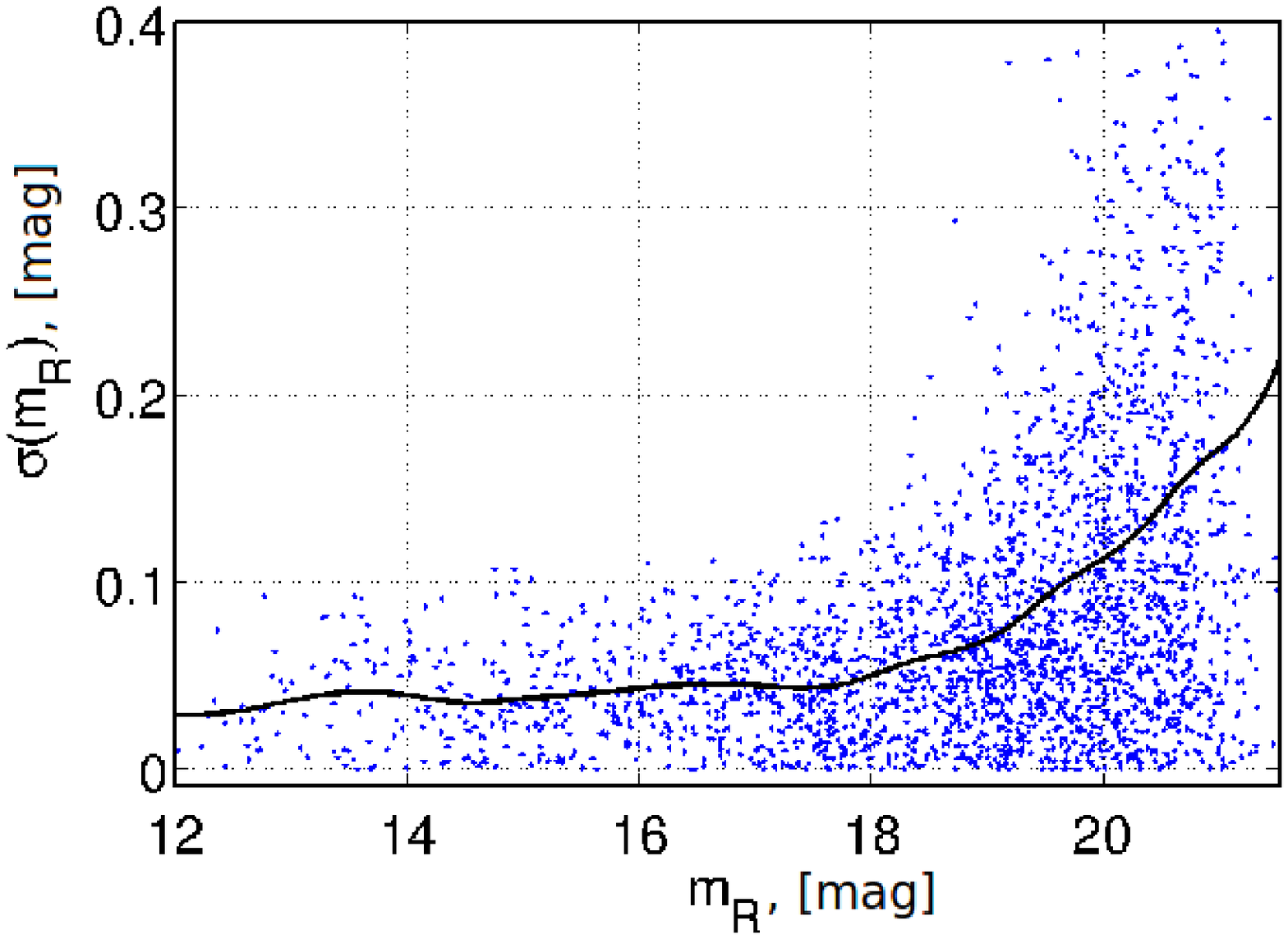} &
(b) & \includegraphics[angle=0,width=70mm]{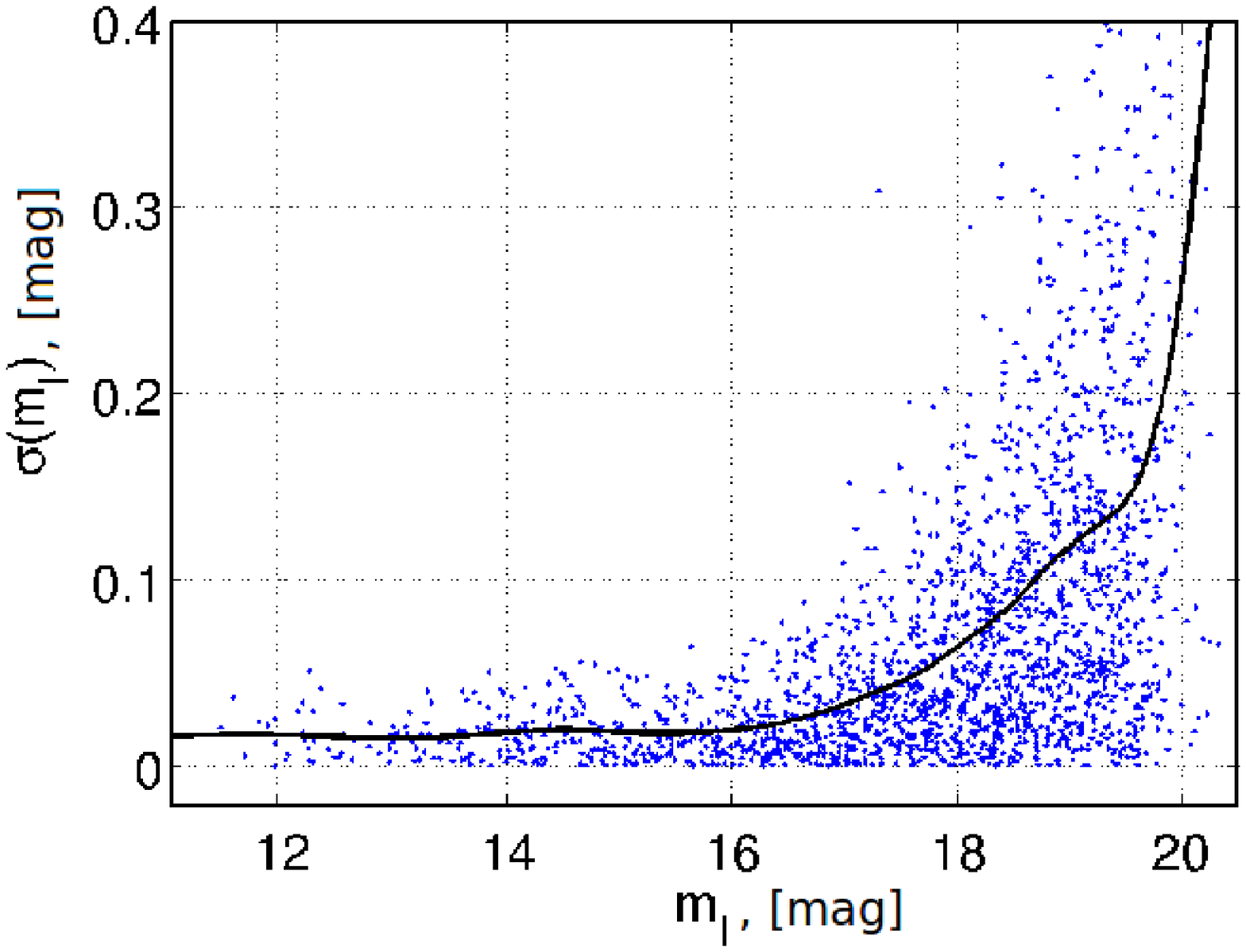} 
\end{tabular}
\end{center}
\caption{Grey magnitude RMS deviation vs. RMS grey magnitude for
$\sim$2200 objects detected more than once. The black
line shows a running median of deviation in 100 object bins. The left panel is for
the m$_R$ magnitudes and the right panel is for the m$_I$ magnitudes.
   \label{Fig:common}}
\end{figure*}

We expect to find multiple detections of the same source during the data reduction process. However, `regular' sources should not change
their brightness significantly and rapidly. Fig. \ref{Fig:common} shows the grey magnitude RMS deviation vs. RMS grey magnitude for
$\sim$2200 objects detected more than once on the images of seven fields from June 13, 2009. The median deviation for both grey 
magnitudes is smaller than the R and I RMS errors for the NCCS limiting magnitudes in Table \ref{Tab:limmag}. A particular source
that changes its magnitude significantly and rapidly is recognized by our pipeline as a `variable' source. 

The final catalogue, after stage two of the photometry, includes two flags defined for each object. The detection flag is 1 when a
particular object has been detected twice - once in R and once in I. Every new detection in R or in I band image
will add one count to this flag. The variability flag is 0 for a non-variable object. Every new detection in R or in I that
will return a magnitude different by $\pm3\sigma$ of the previous detection will change the value of this flag to one, meaning that the object is potentially variable. 

For the final production run the values in the catalogue will be updated following each new detection by adopting weighted means 
and weighted errors of the values. The values defined without errors, such as $\epsilon$(FWHM), will be updated to a simple mean.

\section{Catalogue}

The final catalogue after stage two of photometry includes only entries for objects detected at least once in R and once in I.
The final catalogue does not contain objects defined as truncated by the NCCS pipeline or those with SE internal
flag $>$3. This implies that only objects defined as `regular', or deblended objects, or objects restored 
from the 10\% overlaping with another object, are included in the final catalogue. The final catalogue contains the parameters listed in 
Table \ref{Tab:stage2} and an example of a page from the final catalogue is shown in Table \ref{Tab:fincat}.

\begin{table}[htbp]
\caption{Output Parameters of Final Catalogue:}
\begin{center}
\begin{tabular}{c|c|c}

\hline
\textbf{\#} & \textbf{Catalog Parameter} & \textbf{Units} \\ \hline
1 & J2000.0 $\alpha$ & hh:mm:ss \\ 
2 & $\Delta\alpha$ & sec \\ 
3 & J2000.0 $\delta$ & dd:mm:ss \\ 
4 & $\Delta\delta$ & arcsec \\ 
5 & Kron R flux & ADU/sec \\ 
6 & Kron R flux error & ADU/sec \\ 
7 & Kron I flux & ADU/sec \\ 
8 & Kron I flux error & ADU/sec \\ 
9 & Kron R magnitude & mag \\ 
10 & Kron R magnitude error & mag \\ 
11 & Kron I magnitude & mag \\ 
12 & Kron I magnitude error & mag \\ 
13 & FWHM$_R$ & pix \\ 
14 & FWHM$_I$ & pix \\ 
15 & $a_R$ & pix \\ 
16 & $b_R$ & pix \\ 
17 & $\theta_R$ & deg \\ 
18 & $a_I$ & pix \\ 
19 & $b_I$ & pix \\ 
20 & $\theta_I$ & deg \\ 
21 & $\epsilon$(FWHM) & \textit{unitless} \\ 
22 & $<$FWHM$>$/$<$seeing$>$ & \textit{unitless} \\ 
23 & Point/extended source & \textit{unitless} \\ 
24 & Detection number & \textit{unitless} \\ 
25 & Variability flag & \textit{unitless} \\ \hline
\end{tabular}
\end{center}
\label{Tab:stage2}
\end{table}

\section{Conclusions}

We described procedures, data treatment, and expected photometric and astronomic accuracies of a survey of the NCC performed at the Wise Observatory in the R and I bands with the LAIWO CCD mosaic on the Wise 
Observatory's one meter telescope. The survey detects some 4,000 sources per square degree. The source catalog lists their 
($\alpha, \delta$) coordinates to $\lesssim1''$ and their R and I magnitudes accurate to 0.15 mag or better for sources brighter
than 20.6 in R or 19.6 in I. Essentially $>$90\% of the objects classified by SDSS as point/galactic sources are recognized as such by our survey.
The survey results will be used in conjunction with the data from the TAUVEX UV space telescope to characterize the UV sources.

\section{Acknowledgements}

LAIWO has been built at the Max-Planck-Institute for Astronomy (MPIA) in Heidelberg, Germany with the financial support from 
MPIA, and grants from the German-Israel Foundation and from the Israel Science 
Foundation as a scientific collaboration between Tel Aviv University and MPIA. We are grateful to our German colleagues for 
constructing this instrument and to Dr. Shai Kaspi, the LAIWO liaison scientist 
at Tel Aviv University. We acknowledge the considerable technical help tended by the Wise Observatiry staff, Mr. Ezra Mashal 
and Mr. Sammy Ben Guigui, and MPIA Heidelberg Dr. Karl-Heinz Marien, the project manager, Mr. Ralf Klein, Mr. Florian Briegel, 
and Mr. Harald Baumeister.

Funding for the Sloan Digital Sky Survey (SDSS) has been provided 
by the Alfred P. Sloan Foundation, the Participating Institutions,
the National Aeronautics and Space Administration, the National
Science Foundation, the U.S. Department of Energy, the Japanese
Monbukagakusho, and the Max Planck Society. The SDSS Web site is
http://www.sdss.org/.

The SDSS is managed by the Astrophysical Research Consortium (ARC)
for the Participating Institutions. The Participating Institutions
are The University of Chicago, Fermilab, the Institute for
Advanced Study, the Japan Participation Group, The Johns Hopkins
University, Los Alamos National Laboratory, the
Max-Planck-Institute for Astronomy (MPIA), the
Max-Planck-Institute for Astrophysics (MPA), New Mexico State
University, University of Pittsburgh, Princeton University, the
United States Naval Observatory, and the University of Washington. 

\begin{table}[htbp]
\caption{Final Catalogue Example:}
\begin{center}
\includegraphics[angle=0,width=180mm,angle=90]{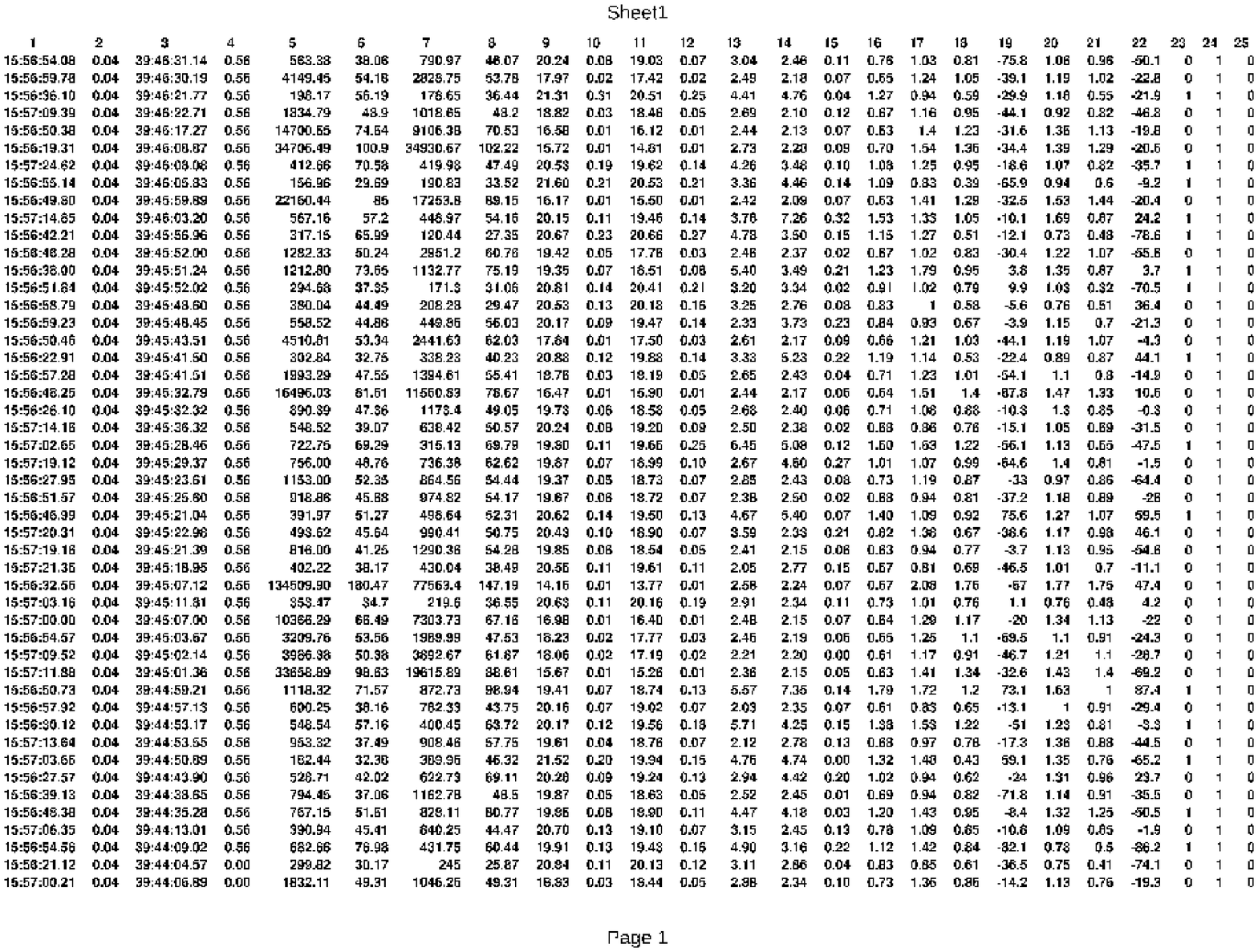}
\end{center}
\label{Tab:fincat}
\end{table}
\clearpage

\end{document}